\documentclass[preprint]{aastex}

\shorttitle{Active M Dwarfs}

\shortauthors{Mochnacki \& et al.}
 
\begin{document}
 
\title{A Spectroscopic Survey of a Sample of Active M Dwarfs.\footnotemark[1]}
\author{Stefan W. Mochnacki\altaffilmark{2}, 
  Michael D. Gladders\altaffilmark{2},
 James R. Thomson, Wenxian Lu\altaffilmark{3},
  Paula Ehlers, Metin Guler,  Asif Hussain,  Quincy Kameda, Karen King, 
Patricia Mitchell, Jason Rowe\altaffilmark{4},
Peter Schindler, \& Heather Scott\altaffilmark{5} }
\affil{David Dunlap Observatory, University of Toronto,
P.O.Box 360, Richmond Hill, Ontario, Canada L4C 4Y6}
\email{stefan@astro.utoronto.ca}

\received{2001 November 29}
\accepted{2002 July 18}
\slugcomment{Accepted 07/18/02 for publication in The Astronomical Journal}
 
\begin{abstract}

A moderate resolution spectroscopic survey of Fleming's sample of 54 X-ray
selected M dwarfs with photometric distances less than 25 pc is presented.
All the objects consist of one or two dMe stars, some being doubles or
spectroscopic binaries. Radial and rotation velocities have been measured by
fits to the H$\alpha$ profiles. Radial velocities have been measured by cross
correlation. Artificial broadening of an observed spectrum has produced a
relationship between H$\alpha$ FWHM and rotation speed, 
which we use to infer rotation speeds for the entire sample by
measurement of the H$\alpha$ emission line.

We find 3 ultra-fast rotators (UFRs, $v \sin i \geq 100 $ km s$^{-1}$), and 8
stars with $30 $ km s$^{-1} \leq v \sin i < 100 $ km s$^{-1}$. 
We find that the UFRs have quite variable emission and should be observed
for photometric variability.
Cross-correlation velocities measured for ultra-fast rotators (UFRs) are
shown to depend on rotation speed and the filtering used. The radial
velocity dispersion of the sample is $17 $ km s$^{-1}$. A new double
emission line spectroscopic binary with a period of 3.55 days has been
discovered, RX~J1547.4+4507, and another known one is in the sample, the
Hyades member RX~J0442.5+2027. Three other objects are suspected
spectroscopic binaries, and at least six are visual doubles.

The only star in the sample observed to have significant lithium happens to
be a known TW Hya Association member, TWA 8A. These results all show that
there are a number of young ($< 10^{8}$ yr) and very young ($< 10^{7}$
yr) low mass stars in the immediate solar neighbourhood.

The H$\alpha$ activity strength does not depend  on rotation speed.
Our fast rotators are  less luminous than similarly fast
rotators in the Pleiades. They are either younger than the Pleiades, or
gained angular momentum in a different way.

\end{abstract}
 

\footnotetext[1]{Based on data obtained at the David Dunlap Observatory,
University of Toronto.}
\altaffiltext{2}{Current address: 
  Observatories of the Carnegie Institution of Washington, 813
  Santa Barbara Street, Pasadena, CA 91101-1292}
\altaffiltext{3}{Current address:
  NASA/GSFC, Mailstop 664, 8800 Greenbelt Road, Greenbelt, MD 20771}
\altaffiltext{4}{Current address: Department of Physics and Astronomy,
  University of British Columbia, 6224 Agricultural Road, Vancouver, B.C.,
  Canada V6T 1Z1}
\altaffiltext{5}{Current address:
  Department of Physics and Astronomy, The University of Western Ontario,
  London, Ontario, Canada N6A 3K7}

\section{INTRODUCTION}

New technologies and space astronomy have led to the accumulation of
catalogs of objects observed in widely different parts of the
electromagnetic spectrum. Cross-correlating these catalogs has become a
fruitful pursuit. \citet{f98} published photometry of a sample of 54 M
dwarf stars which were selected on the basis of detection in X-rays as part
of the ROSAT All-Sky Survey and being red in photographic sky surveys, and
which had apparent photometric parallaxes placing them closer than 25
parsecs from the Sun. Few of these stars had been studied spectroscopically,
so we set out to observe their H$\alpha$ and Li lines and to measure their
radial velocities. We wanted to see whether their X-ray brightness indicated
strong chromospheric activity: what fraction of Fleming's stars are dMe
stars?

\citet{f98} had shown that these stars probably have small proper
motions, so we were interested in seeing whether other indications of youth
are present. It has long been known that single M dwarfs decline in
activity with age. The discovery of the TW Hydrae Association
\citep{kz97,wz99} shows that very young stars can be found in the immediate
solar neighbourhood. Not only H$\alpha$ emission, but also Li absorption and
rotation as well as space motion are used as diagnostics of youth.

By taking spectra over some period of time, we hoped to be able to find
spectroscopic binaries among these stars and to estimate their binary
frequency. \citet{fm92} have done this thoroughly for 
well-defined samples of relatively bright M dwarfs, including a few dMe
stars. More recently,  the discovery of brown dwarfs has made searching
for companions to low-mass stars a desirable goal, as reviewed by
\citet{b00}.

\section{OBSERVATIONS}

\subsection{Sample}

Stars apparently closer than 25 pc in the survey by \citet{f98} were
observed spectroscopically with the Cassegrain spectrograph on the David
Dunlap Observatory 1.88m telescope. The combination of a 306$\mu$ slit
subtending 1.85\arcsec~ on the sky, 1800 l/mm grating and front-illuminated
Thomson-CSF THX31156 detector produced a spectral purity (inverse
resolution) of 0.45 \AA~ with an instrumental profile of full width at half
maximum (FWHM) of 3.0 pixels. The spectrograph was focused at the beginning
of each night, with allowance for expected temperature changes. The image of
the 30\arcsec~ slit was binned in 50 columns, corresponding to 200 true
columns on the CCD. There was no binning of the 1024 pixels in the direction
of dispersion. Exposures were usually of 30 minutes duration. Objects were
easy to find thanks to the on-line availability of the Digitized Sky Survey,
an intensified VARO-EEV CCD guiding camera and consistent digital telescope
positioning. The number of observations per object varied widely, due to the
uneven distribution of objects around the sky, the vagaries of weather and
because some objects were quickly found to be unusual. The approximately
200~\AA~ wavelength range allowed both H$\alpha$ and the Li $\lambda$ 6708
feature to be observed simultaneously.

The observing program was begun 1998 September, as an undergraduate class
project. We quickly found that all objects in Fleming's sample had
well-resolved H$\alpha$ in emission. Two double-dMe spectroscopic binaries
were found, and it was also clear that several stars had very broad
H$\alpha$ emission. The continua of these stars were not so obvious due to
the high sky background at the DDO. Further observations were undertaken at
various intervals until 2001 April.

The seeing at the DDO was frequently in the range of 1-2\arcsec, and the
intensified guider allowed us to resolve several close visual binaries. The
most southerly object in this sample, RX~J1132.7-2651, is a visual double
dMe, and as we shall see, it is extremely interesting. Both components of
all of the doubles are dMe, except for RX~J0324.1+2347A, which has H$\alpha$
in absorption, while the companion of RX~J1509.0+5904 was too faint to be
observed ( $V \backsimeq 16$ was our spectroscopic limit in practice). The
photometry of \citet{f98} probably included both components
of each double.

M dwarfs taken from the list of \citet{mlw87} were
observed each night as standards. Three of these stars were chosen as
templates for determining radial velocities.
 
\subsection{Reductions}

After initial reductions and analysis, all spectra were re-reduced by SWM
with an IRAF script developed for this survey, based on the techniques used
by
\citet{gc98}. Care was taken to remove cosmic rays, to which the detector
is quite susceptible, and which are made more difficult to remove by the
binning perpendicular to the direction of dispersion. Flattening was also an
issue due to  non-uniform illumination of the slit by the comparison
lamps. Given the high sky brightness at the DDO, the subtraction of
backround and the extraction of stellar continua demanded considerable
interactive control of IRAF procedures. 

\begin{figure}
\figurenum{1}
\plotone{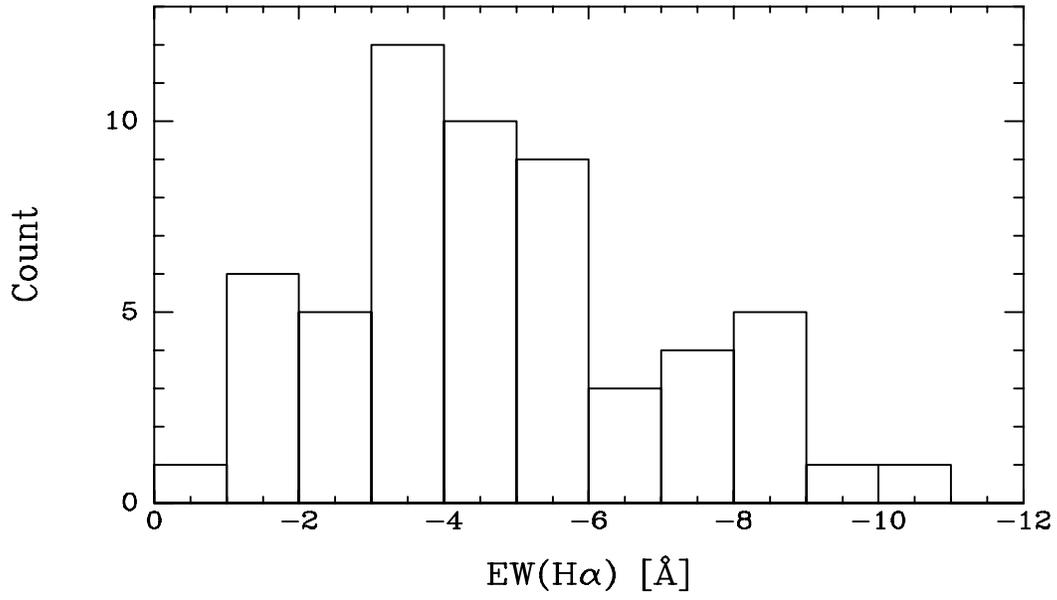}
\figcaption[fig1.eps] {\label{fig1}
Distribution of average equivalent widths of the H$\alpha$ emission.
}
\end{figure}

\subsection{H$\alpha$ emission line measurements}

The H$\alpha$ lines were measured using the ``splot" task in IRAF. Single or
double Gaussians were fitted, with the continuum chosen in the region of
6555 - 6575 \AA. Since the equivalent width of the H$\alpha$ emission is
typically in the range -1 to -10 \AA~ for these stars (negative equivalent
widths indicating emission), these fits could be consistently done even when
the continuum was too weak for cross-correlation. The fits produce a
self-consistent set of radial velocities, widths and equivalent widths,
which appear in the right half of Table~\ref{tab1}, while the average
H$\alpha$ measurements are in the right half of Table~\ref{tab2}. The
distribution of average H$\alpha$ emission equivalent widths is shown in
Figure~\ref{fig1}.

 
 

\begin{deluxetable}{llrrrrrrrcc}
\label{tab1}
\pagestyle{empty}
\tabletypesize{\scriptsize}
\setlength{\tabcolsep}{0.03in}
\tablecolumns{11}
\tablewidth{500pt}
\tablenum{1}
\tablecaption{DDO Observations of Active M Dwarfs\label{tab1}}
\tablehead{ 
\colhead{Object} &	
\colhead{File} &	
\colhead{Time} &	
\colhead{$W_{fx}$} &	
\colhead{$V_{fx}$} &	
\colhead{$Err(V_{fx})$} & 
\colhead{$EW(H\alpha)$} & 
\colhead{$W_{H\alpha}$}  & 
\colhead{$V_{H\alpha}$} & 
\colhead{FC} &		
\colhead{SBC} \\	
\colhead{(RASS Desig.)} & 
\colhead{(name)} &	
\colhead{(HJD)} &	
\colhead{(km s$^{-1}$)} &   
\colhead{(km s$^{-1}$)} &   
\colhead{(km s$^{-1}$)} &   
\colhead{(\AA~)} &      
\colhead{(\AA~)} &      
\colhead{(km s$^{-1}$)} &  
\colhead{} &		
\colhead{} \\		
\colhead{(1)} &
\colhead{(2)} &
\colhead{(3)} &
\colhead{(4)} &
\colhead{(5)} &
\colhead{(6)} &
\colhead{(7)} &
\colhead{(8)} &
\colhead{(9)} &
\colhead{(10)} &
\colhead{(11)}
}

\startdata
RX~J0016.9+2003  & F010120\_6     & 2451929.5730 &  33.3 &   6.9 &  5.4 &  \nodata & \nodata &  \nodata & 1 & \nodata \\ 
RX~J0016.9+2003  & F981129\_3     & 2451146.5248 &  36.0 &  -4.1 &  8.5 &  -2.3 & 1.47 &   2.0 & 1 & \nodata \\ 
RX~J0016.9+2003  & F981227\_2     & 2451174.4740 &  29.4 &   1.4 &  2.7 &  -2.6 & 1.39 &  -1.4 & 1 & \nodata \\ 
RX~J0019.7+1951  & F981010\_8     & 2451096.7102 &  30.6 &  -5.5 &  6.3 &  -6.4 & 1.48 &  -8.4 & 1 & \nodata \\ 
RX~J0019.7+1951  & F981010\_8p9   & 2451096.7102 &  42.0 &  -9.7 &  9.3 &  \nodata & \nodata &  \nodata & 1 & \nodata
\tablecomments{The complete version of this table is in the electronic
edition of the Journal.  The printed edition contains only a sample.}
\enddata
\end{deluxetable}
 

A more physically meaningful way of considering H$\alpha$ emission is to
compute its {\it flux} and to compare it with the bolometric luminosity of
each star. This is demonstrated, for example, in the case of the Hyades by
\citet{rm00}, and discussed in Chapter 5 of \citet{rh00}.

We first need to compute the continuum flux $F_{C}$ at H$\alpha$ to convert
equivalent widths into fluxes. \citet{rhm95} show that this is given by,

\begin{equation}
F_{C} = 1.45 \frac{F_{0R}}{\lambda_{R}^{2}} 10^{-0.4 M_{R}}  
\end{equation}
 
\noindent
where the flux zero-point $F_{0R}$ in R is $2750 Jy$ \citep{rg84}. The
absolute magnitude $M_{V}$ as a function of $V-I$ is taken from
\citet{spi89}, while  $V-R$ as a function of $V-I$ for M dwarfs was derived
by \citet{hgr96}. The bolometric correction $BC_{I}$ was derived by
\citet{lab96} and used by \citet{rm00}. Applying these calibrations to the
photometry of \citet{f98} and our measurements produces values of
$L_{H\alpha}$ and $L_{bol}$, as shown in Table~\ref{tab3}. 

In the case of spectroscopic binaries, the equivalent widths of the  H$\alpha$ emission
from each component need to be corrected due to the continuum of the other
component. We assume, for lack of any information better than a mass ratio
of unity for RX~J1547.4+4507, that the components in each binary have equal
luminosity. The average equivalent widths in Table 2 are corrected
accordingly: the measured and averaged equivalent widths of the resolved
components are multiplied by 2.0 to allow for the continuum contamination of
the other component. In the case of RX~J0442.5+2027, we have not been able to
identify the components, and the equivalent widths appear to be similar.
The same correction is made in Tables 2 and 3. No other stars are adjusted:
identified or suspected double stars are noted, since the published
photometry for these stars is affected by their multiplicity.

\subsection{Absorption line system radial velocities}

The ``fxcor" task in IRAF was used to measure the
absorption line system radial velocities of all program and standard stars.
Three well-observed M dwarfs were chosen as templates: HD36395 (Gliese 205),
HD119850 (Gliese 526) and HD95735 (Gliese 411). Spectra of each of these
stars taken 2001 Jan 3 were chosen because that night was exceptionally dry
and cold (-16 C at the telescope secondary mirror), with unusually little
telluric absorption. HD87901 (Regulus) was observed as the telluric
standard. The spectrum was carefully fitted with high-order polynomials to
remove the broad absorption features of Regulus, and then the ``telluric"
task of IRAF was applied to the radial velocity template spectra. The
processed template spectra were further fitted with polynomials and divided
to remove low-frequency variations. Telluric correction of program stars was
not undertaken, since the high S/N template spectra were already well
corrected and therefore there should be no telluric component in the
cross-correlations.

One difficulty was detected after all analysis had been done: in some cases
the velocity used for the template spectrum of HD95735 was in error by
$-3.00$ km s$^{-1}$.  Since all velocities are averages of the
cross-correlation velocities using all three templates, some of our
absorption velocities may have a systematic error of about $-1 $ km
s$^{-1}$. This is less than any standard deviation of relevance here, but
may need to be taken into account when combining our results with others.

The three template stars were also chosen because they span populations
(HD95735 is a high-velocity star), though they are not as late as many of
our dMe program stars. The H$\alpha$ region (6555-6570 \AA) was not used in
the cross-correlations. Considerable effort was invested into finding
Fourier filtering parameters which produced good cross-correlation profiles.
The ``Welch" filtering function was found to work best, with cuton and
cutoff values of 4 and 400 being used typically (the Nyquist frequency being
512 in the digital Fourier domain). Slow and moderate rotators produced good
cross-correlation profiles, but weakly exposed spectra did not.
Cross-correlations with formal uncertainties of more than $10 $ km s$^{-1}$
have been generally discarded, but sometimes smaller uncertainties were
associated with clearly poor fits.




\begin{deluxetable}{lrrrrrrrrccrrrrr}
\label{tab2}
\rotate
\tabletypesize{\scriptsize}
\pagestyle{empty}
\setlength{\tabcolsep}{0.03in}
\tablecolumns{16}
\tablewidth{0pc}
\tablenum{2}
\tablecaption{Averaged Spectroscopic Data of Active M Dwarf
Sample\label{tab2}}
\tablehead{
\colhead{Object} &  				
\colhead{N$_{fx}$} & 				
\colhead{$W_{fx}$} &				
\colhead{$\sigma(W_{fx})$} &			
\colhead{$V_{fx}$} &				
\colhead{$\sigma(V_{fx})$} &			
\colhead{$\overline{Err(V_{fx})}$} &		
\colhead{$V_{fx}(unb)$} &			
\colhead{N$_{H\alpha}$} &			
\colhead{$EW(H\alpha)$} &			
\colhead{$\sigma(EW_{H\alpha})$} &		
\colhead{$W_{H\alpha}$}  &			
\colhead{$\sigma(W_{H\alpha})$} &		
\colhead{$V_{H\alpha}$} &			
\colhead{$\sigma(V_{H\alpha})$} &		
\colhead{$v \sin i$} \\				
\colhead{(RASS Desig.)} &				
\colhead{} &					
\colhead{(km s$^{-1}$)} &				
\colhead{(km s$^{-1}$)} & 			
\colhead{(km s$^{-1}$)} &				
\colhead{(km s$^{-1}$)} &				
\colhead{(km s$^{-1}$)} &				
\colhead{(km s$^{-1}$)} &				
\colhead{} &					
\colhead{(\AA~)} &				
\colhead{(\AA~)} &				
\colhead{(\AA~)} & 				
\colhead{(\AA~)} & 				
\colhead{(km s$^{-1}$)} & 			
\colhead{(km s$^{-1}$)} & 			
\colhead{(km s$^{-1}$)} \\			
\colhead{(1)} &
\colhead{(2)} &
\colhead{(3)} &
\colhead{(4)} &
\colhead{(5)} &
\colhead{(6)} &
\colhead{(7)} &
\colhead{(8)} &
\colhead{(9)} &
\colhead{(10)} &
\colhead{(11)} &
\colhead{(12)} &
\colhead{(13)} &
\colhead{(14)} &
\colhead{(15)} &
\colhead{(16)}
}

\startdata
RX~J0016.9+2003  & 3 &  32.9 &   3.3 &   1.4 &    5.5 &    5.5 &     1.4 &     2 &   -2.5 &    0.2 &   1.43 &      0.06 &   0.3 &     2.4 &    22  \\  
RX~J0019.7+1951  & 3 &  37.7 &   6.3 &  -9.6 &    4.0 &    8.8 &    -9.6 &     3 &   -5.7 &    0.7 &   1.36 &      0.15 &  -7.2 &     1.0 &   $<$20  \\  
RX~J0024.5+3002  & 2 &  41.7 &   0.6 &   9.5 &    1.7 &    7.0 &     9.5 &     3 &   -6.4 &    1.7 &   1.29 &      0.03 &   9.2 &     1.0 &   $<$20  \\  
RX~J0048.9+4435  & 2 &  41.4 &   9.8 & -14.8 &    0.6 &    5.8 &   -14.8 &     2 &   -8.2 &    5.1 &   1.69 &      0.39 & -18.4 &     1.8 &    36  \\  
RX~J0050.5+2449  & 3 &  34.2 &   7.3 &   6.4 &    0.5 &    3.0 &     6.4 &     3 &   -4.7 &    0.3 &   1.39 &      0.01 &   5.6 &     1.2 &   $<$20  \\  
RX~J0102.4+4101  & 4 &  37.6 &  10.4 &   9.2 &   37.0 &    7.0 &     7.6 &     3 &   -2.0 &    0.1 &   2.31 &      0.02 &  -3.7 &     2.9 &    62  \\  
RX~J0111.4+1526  & 3 &  38.3 &   2.8 &   4.5 &    6.7 &    6.4 &     4.5 &     3 &   -7.0 &    0.7 &   1.40 &      0.01 &   2.8 &     1.9 &    20  \\  
RX~J0122.1+2209  & 3 &  45.1 &  13.0 &  -0.7 &    2.7 &    6.6 &    -0.7 &     3 &   -4.1 &    0.7 &   1.40 &      0.02 &  -3.8 &     1.0 &    20  \\  
RX~J0123.4+1638  & 1 &  48.3 & \nodata &   2.9 &  \nodata &  7.5 &  2.9 &  1 &  -5.5 &  \nodata &  1.41 &  \nodata &   9.4 &   \nodata &   21   \\ 
RX~J0143.1+2101  & 1 &  31.5 & \nodata & -47.0 & \nodata &   2.7 &  -47.0 &  1 &  -3.2 & \nodata &   1.21 & \nodata & -45.5 & \nodata &  $<$20   \\ 
RX~J0143.6+1915  & 2 & 102.5 &   3.1 &  10.0 &    4.1 &   17.8 &     0.6 &     2 &   -8.8 &    1.1 &   4.01 &      0.04 &  -7.4 &    17.0 &   121  \\  
RX~J0212.9+0000  & 4 &  33.2 &   4.1 &  27.5 &    1.8 &    3.8 &    27.5 &     4 &   -2.2 &    0.2 &   1.28 &      0.09 &  26.1 &     0.7 &   $<$20  \\  
RX~J0219.0+2352  & 3 &  36.3 &   2.7 &  16.3 &    4.6 &    5.8 &    16.3 &     2 &   -8.8 &    1.3 &   1.47 &      0.04 &  15.6 &     3.4 &    25  \\  
RX~J0249.9+3345A & 1 & 134.5 & \nodata &  30.7 & \nodata &  29.9 &  30.7 &     1 &   -8.6 & \nodata &   1.96 & \nodata &  4.7 & \nodata &   48   \\ 
RX~J0249.9+3345B & 1 & 33.0 & \nodata &  7.7 & \nodata &    3.9 &     7.7 &     1 &   -7.7 & \nodata &   1.66 & \nodata &  -1.3 & \nodata &   35   \\ 
RX~J0324.1+2347A & 4 &  35.8 &   6.2 &  19.0 &    1.6 &    2.2 &    19.0 &     0 & \nodata & \nodata & \nodata & \nodata & \nodata & \nodata & \nodata  \\  
RX~J0324.1+2347B & 3 &  36.6 &   4.6 &  25.1 &   15.4 &    3.2 &    25.1 &     3 &   -3.5 &    0.8 &   1.33 &      0.06 &  25.0 &    16.4 &   $<$20  \\  
RX~J0332.6+2843  & 5 &  41.7 &   5.5 &   7.4 &    3.1 &    5.4 &     7.4 &     5 &   -5.8 &    1.6 &   1.47 &      0.10 &   7.5 &     1.3 &    25  \\  
RX~J0339.4+2457  & 4 &  29.8 &   1.6 &  12.3 &    1.5 &    2.4 &    12.3 &     4 &   -2.4 &    1.3 &   1.61 &      0.28 &   8.9 &     3.9 &    32  \\  
RX~J0349.7+2419  & 3 &  37.5 &   3.6 &  29.3 &    5.9 &    7.5 &    29.3 &     2 &   -5.7 &    0.1 &   1.36 &      0.08 &  31.4 &     0.2 &   $<$20  \\  
RX~J0442.5+2027A,B  & 52 &  32.9 &   5.1 &  41.4 &   32.6 &    3.9 &    41.4 & 50 &   -3.8\tablenotemark{a} &    0.8 &   1.38 &      0.11 &  39.6 &    32.9 &   $<$20  \\  
RX~J0446.1+0644  & 6 &  37.0 &   2.3 &  43.5 &    2.0 &    6.2 &    43.5 &     7 &   -4.2 &    0.4 &   1.25 &      0.13 &  43.0 &     3.8 &   $<$20  \\  
RX~J0448.7+1003  & 8 &  31.8 &   1.1 &  10.6 &    1.7 &    1.9 &    10.6 &     8 &   -1.0 &    0.2 &   1.44 &      0.15 &   2.2 &     9.3 &    23  \\  
RX~J0747.2+2957  & 15 &  38.5 &   3.6 &   7.9 &    2.9 &    4.1 &     7.9 &    15 &   -3.4 &    0.5 &   1.50 &      0.08 &   6.1 &     2.8 &    26  \\  
RX~J1002.8+4827  & 13 &  37.4 &   4.8 &  -3.3 &    4.0 &    6.7 &    -3.3 &    15 &   -7.4 &    0.9 &   1.43 &      0.04 &  -3.3 &     1.6 &    22  \\  
RX~J1038.4+4831  & 15 &  31.6 &   3.1 &  15.0 &    1.6 &    2.7 &    15.0 &    15 &   -4.9 &    1.0 &   1.32 &      0.12 &  14.9 &     1.1 &   $<$20  \\  
RX~J1132.7-2651A & 7 &  32.6 &   5.1 &   7.3 &    1.3 &    3.2 &     7.3 &     7 &   -8.8 &    3.2 &   1.62 &      0.17 &   7.4 &     1.2 &    33  \\  
RX~J1132.7-2651B & 1 &  37.4 & \nodata &  -1.2 & \nodata &    7.7 &    -1.2 &     3 &  -10.8 &    2.2 &   1.47 &    0.14 &  6.4 &  1.4 &  25  \\  
RX~J1221.4+3038A & 14 &  35.4 &   7.5 &   4.4 &    4.7 &    4.8 &     4.4 &    16 &   -3.3 &    0.6 &   1.18 &      0.07 &   3.5 &     3.2 &   $<$20  \\  
RX~J1221.4+3038B & 13 &  35.1 &   7.5 &   4.9 &    1.8 &    4.6 &     4.9 &    15 &   -3.9 &    1.5 &   1.21 &      0.19 &   3.6 &     2.3 &   $<$20  \\  
RX~J1310.1+4745  & 7 &  29.6 &   2.2 & -13.5 &    2.1 &    3.2 &   -13.5 &     7 &   -3.4 &    0.8 &   1.23 &      0.07 & -13.9 &     1.1 &   $<$20  \\  
RX~J1332.6+3059  & 6 &  46.0 &   7.1 & -15.4 &    7.3 &    8.3 &   -15.4 &     6 &   -4.5 &    0.6 &   1.56 &      0.07 & -15.1 &     1.0 &    30  \\  
RX~J1348.7+0406  & 4 &  35.5 &   6.3 &  -0.0 &    1.0 &    4.8 &    -0.0 &     5 &   -3.7 &    0.6 &   1.35 &      0.07 &  -0.2 &     2.2 &   $<$20  \\  
RX~J1351.8+1247  & 5 &  30.4 &   2.6 &  -6.6 &    1.8 &    1.7 &    -6.6 &     5 &   -2.2 &    0.3 &   1.34 &      0.03 &  -7.6 &     0.4 &   $<$20  \\  
RX~J1359.0-0109  & 5 &  33.2 &   3.2 & -17.1 &    0.7 &    3.7 &   -17.1 &     5 &   -4.1 &    0.3 &   1.34 &      0.07 & -17.6 &     0.9 &   $<$20  \\  
RX~J1410.9+0751  & 5 &  29.7 &   2.4 &  -1.6 &    1.2 &    2.1 &    -1.6 &     5 &   -3.6 &    0.6 &   1.31 &      0.05 &  -2.5 &     0.7 &   $<$20  \\  
RX~J1420.0+3902  & 13 & 133.7 &  23.1 & -12.7 &    5.2 &   15.5 &   -20.6 &    13 &   -5.0 &    0.7 &   3.66 &      0.18 & -24.6 &     8.4 &   109  \\  
RX~J1432.1+1600  & 4 &  31.9 &   1.3 &  -3.4 &    0.5 &    2.7 &    -3.4 &     4 &   -4.8 &    1.1 &   1.41 &      0.08 &  -3.1 &     2.7 &    21  \\  
RX~J1438.7-0257  & 4 &  34.3 &   2.1 & -27.0 &    1.4 &    5.0 &   -27.0 &     4 &   -3.4 &    0.5 &   1.39 &      0.15 & -27.0 &     4.1 &   $<$20  \\  
RX~J1447.2+5701  & 6 &  34.9 &   4.6 &  -4.2 &    3.1 &    4.2 &    -4.2 &     5 &   -5.8 &    0.4 &   1.36 &      0.03 &  -5.3 &     2.3 &   $<$20  \\  
RX~J1459.4+3618  & 5 &  33.1 &   2.7 & -22.8 &    1.4 &    3.9 &   -22.8 &     5 &   -6.9 &    0.6 &   1.25 &      0.01 & -21.4 &     1.0 &   $<$20  \\  
RX~J1509.0+5904  & 7 &  32.7 &   1.9 &  -5.4 &    3.3 &    3.3 &    -5.4 &     6 &   -3.3 &    0.5 &   1.34 &      0.06 &  -4.9 &     1.4 &   $<$20  \\  
RX~J1512.6+4543  & 7 &  35.8 &   6.4 &  -8.3 &    7.3 &    4.4 &    -8.3 &     7 &   -4.0 &    0.4 &   1.41 &      0.06 & -12.9 &     1.8 &    21  \\  
RX~J1523.8+5827  & 3 &  36.0 &   3.3 & -15.8 &    1.4 &    3.4 &   -15.8 &     5 &   -5.9 &    1.4 &   1.32 &      0.05 & -13.6 &     3.1 &   $<$20  \\  
RX~J1529.0+4646  & 2 &  33.7 &   2.0 &   8.5 &    0.3 &    5.1 &     8.5 &     3 &   -3.2 &    0.5 &   1.17 &      0.13 &   6.4 &     0.8 &   $<$20  \\  
RX~J1542.3+5936  & 7 &  36.3 &   5.5 & -20.2 &    2.4 &    5.7 &   -20.2 &     7 &   -7.5 &    1.5 &   1.25 &      0.15 & -21.3 &     0.9 &   $<$20  \\  
RX~J1547.4+4507A & 42 & 33.5 &   5.2 & -21.4 & 0.4    &    4.0 &   -21.4 &   48  &   -2.5\tablenotemark{a} &    1.0  &   1.25 &     0.11 & -21.4 &    0.4 &   $<$20  \\  
RX~J1547.4+4507B & 42 & 33.6 &   5.2 & -21.4 & 0.4    &    4.2 &   -21.4 &   48  &   -3.8\tablenotemark{a} &    0.8  &   1.25 &     0.11 & -21.4 &    0.4 &   $<$20  \\ 
RX~J1548.0+0421  & 3 &  32.4 &   0.7 &  -8.5 &    2.6 &    3.1 &    -8.5 &     3 &   -3.1 &    0.3 &   1.31 &      0.03 &  -9.6 &     1.4 &   $<$20  \\  
RX~J1648.0+4522  & 7 &  31.6 &   2.9 & -14.5 &    1.5 &    3.6 &   -14.5 &     7 &   -5.4 &    1.2 &   1.24 &      0.05 & -14.3 &     2.0 &   $<$20  \\  
RX~J2137.6+0137  & 3 & 113.2 &  27.6 & -12.2 &    3.0 &   25.2 &   -12.8 &     4 &   -9.2 &    1.4 &   2.13 &      0.19 &  -8.8 &     2.9 &    55  \\  
RX~J2227.8-0113  & 6 & 122.3 &  20.3 &  -1.3 &   19.6 &   25.0 &    -9.6 &     8 &   -6.9 &    1.3 &   3.75 &      0.43 & -25.1 &     7.3 &   112  \\  
RX~J2243.7+1916  & 5 &  34.6 &   2.7 &  13.5 &    3.2 &    4.2 &    13.5 &     4 &   -1.1 &    0.2 &   1.33 &      0.14 &  10.9 &     3.6 &   $<$20  \\  
RX~J2317.5+3700  & 4 &  38.4 &   6.7 &   3.8 &    2.0 &    2.9 &     3.8 &     3 &   -0.6 &    0.1 &   1.12 &      0.12 &  26.7 &     1.2 &   $<$20  \\  
RX~J2326.2+2752  & 2 &  38.9 &   4.4 &  10.1 &    2.1 &    5.4 &    10.1 &     2 &   -5.2 &    1.3 &   1.46 &      0.19 &  13.5 &    10.1 &    24  \\  
RX~J2333.3+2714  & 6 &  36.5 &   5.3 &  -0.3 &    2.1 &    5.3 &    -0.3 &     6 &   -1.7 &    0.2 &   1.32 &      0.12 &  -3.3 &     4.7 &   $<$20  \\  
RX~J2337.5+1622  & 0 & \nodata & \nodata & \nodata & \nodata & \nodata & \nodata &     1 &   -4.7 & \nodata &   1.67 & \nodata &   1.1 & \nodata & 35  \\  
RX~J2349.2+1005  & 4 &  31.8 &   3.9 &  26.2 &   21.7 &    5.1 &    26.2 &     3 &   -1.8 &    0.2 &   1.59 &      0.37 &  27.4 &    18.1 &    31  \\  
RX~J2354.8+3831  & 3 &  30.6 &   3.0 &   5.0 &    1.2 &    3.3 &     5.0 &     2 &   -4.7 &    0.1 &   1.26 &      0.03 &   5.3 &     0.5 &   $<$20 
\tablenotetext{a}{Corrected for binarity}
\enddata
\end{deluxetable}


The ultra-fast rotators (UFRs) produced very poor cross-correlation
profiles; better results were obtained with Welch filter frequency range
(6,200), for 1024-pixel one-dimensional spectra. These results are discussed
below.

For the double-lined spectroscopic binaries, the absorption
cross-correlation results gave somewhat better resolution than the H$\alpha$
measurements. This is because the ``fxcor" FWHM in our data was typically
$30-35 $ km s$^{-1}$ (i.e. $\sqrt{2} \times$ the spectral purity), whereas the
resolved width of a slowly-rotating dMe H$\alpha$ profile was about 1.25
\AA~, or $56 $ km s$^{-1}$. For the ultra fast rotators, however, the more
restricted Fourier frequency range was equivalent to smoothing the spectra,
and the cross-correlation profiles were wider to begin with.

\begin{figure}
\figurenum{2}
\plotone{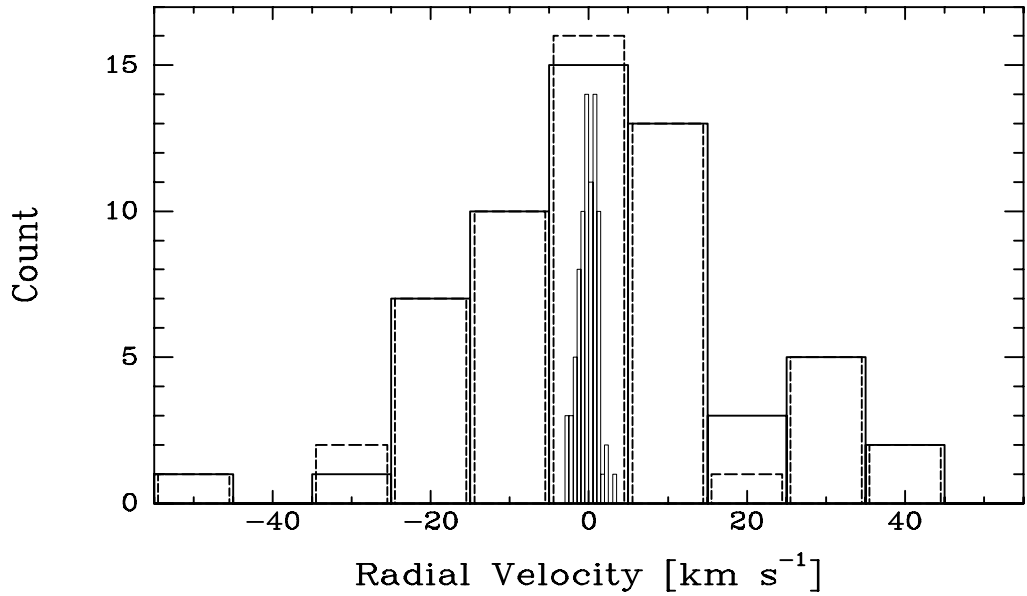}
\figcaption[fig2.eps]{\label{fig2}
Distribution of radial velocities. The large solid line bars represent
cross-correlation (absorption line system) velocities of the program stars,
and the dashed line bars represent H$\alpha$ emission line velocities. The
narrow bins in the centre, $0.25 $ km s$^{-1}$ wide, represent the
distribution of the differences between our velocities and the published
velocities of M dwarf standards \citep{fm92}.}
\end{figure}

Figure~\ref{fig2}  contains a histogram of all our M dwarf radial velocity
standard observations with their published velocities \citep{mlw87}
subtracted. The standard deviation, with no corrections applied other than
the standard heliocentric ones, is $1.2 $ km s$^{-1}$, and the formal mean is
$-0.1 $ km s$^{-1}$. 

Our observations are listed in Table 1.  Column 1 is the full RASS
designation of each star. We note that \citet{f98} did not publish the full
designations; the SIMBAD resolver requires the names as we present them,
with the blank between RX and J, and the fourth declination digit. We
add A or B to designate double components of visual doubles, where separate
spectra were taken. Column 2 is the file name under which the reduced
spectra are archived at \url{http://www.astro.utoronto.ca}. Column 3 is
the heliocentric Julian date at mid-observation. Column 4 is the full
width at half-maximum of the cross-correlation profile, and column 5
contains the error estimate computed by the ``fxcor" task of IRAF. The
H$\alpha$ equivalent width, FWHM and velocity, from fitting a Gaussian, are
in columns 7, 8 and 9 respectively. Column 10, labelled ``FC", indicates
which fitted profile, from blue to red, is listed, while column 11,
labelled ``SBC", shows to which component of the spectroscopic binary the
measured profile is assigned. Only in the case of RX~J1547.4+4507 were we
able to assign lines to binary components.

The individual velocity and line measurements for each star have been
averaged and are reproduced in Table~\ref{tab2}. Column 2 is the number of
cross-correlations used, column 3 the FWHM of the cross-correlation
profile and column 4 shows the standard deviation of the cross-correlation
FWHM (useful to detect variability). Column 5 contains the ``fxcor"
velocities, while column 6 is the root mean square deviation of the
``fxcor" velocities and column 7 has the {\it average} of the individual
errors computed by ``fxcor" (also useful for detecting variability). Column
8 has velocities with the predicted bias subtracted for $v \sin i > 50 $ km
s$^{-1}$.

The right hand side of Table 2 contains H$\alpha$ measurements. Column 9
is the number of spectra of each star. Differences  between the number of
cross-correlation measurements and H$\alpha$ measurements are due to
difficulties with one or the other, such as too weak a continuum exposure, or
bad wavelength calibration at H$\alpha$ due to the grating position being
incorrect. Columns 10 and 11  show the average H$\alpha$ equivalent width
and its standard deviation, respectively, while columns 12 and 13
contain mean FWHM values and their standard deviations. Columns 14 and
15 show mean H$\alpha$ radial velocities and their standard deviations,
while the final column shows the rotation speed derived from the FWHM of
H$\alpha$.

The distributions of H$\alpha$ and ``fxcor" velocities averaged for each
star are also shown in Figure~\ref{fig2}.

\subsection{Rotation velocity measurements}
 
To estimate the projected rotation speeds $v \sin i$, we have calibrated a
relationship between the FWHM of Gaussians fitted to the H$\alpha$ profiles of
spectra and the rotation speed of an artificially broadened dMe spectrum
with the same fitted Gaussian FWHM. The summed, velocity-shifted spectrum of
RX~J1038+4831 was chosen as the spectrum to be artificially convolved with the
rotational broadening function,

\begin{equation}
G(\Delta\lambda) = \frac{2(1-\epsilon)[1-(\Delta\lambda/\Delta\lambda_{L})^{2}
]^{\case{1}{2}} + \frac{1}{2}\pi\epsilon[1-(\Delta\lambda/
\Delta\lambda_{L})^{2}]}{\pi\Delta\lambda_{L}(1-\epsilon/3)} ,
\end{equation}

\noindent
as given in the text by \citet{g92}, where $\epsilon$ is the linear
limb-darkening coefficient at the observed wavelength and
$\Delta\lambda_{L}$ is the wavelength shift at the limb of the star,
corresponding to the projected equatorial rotation velocity $v \sin i$. This
star was well observed, reasonably bright, and not a radial velocity
variable. Its H$\alpha$ emission line is relatively narrow and strong, and
its colours are reasonably similar to some of the ultrafast rotators. This
spectrum was artificially broadened at $10 $ km s$^{-1}$ intervals up to 
$v\sin i = 150$ km s$^{-1}$, for limb-darkening constants $\epsilon$ of 0.8
and 0.

The broadened spectra were measured in exactly the same way as all the
spectra in the sample. The H$\alpha$ measurements of the spectra broadened
with zero limb darkening are used as our calibration for estimating $v \sin i$
from the full width at half maximum of Gaussian fits. Although Gaussians
do not fit resolved dMe H$\alpha$ profiles perfectly \citep{sb97},
the fits are done consistently and there is a fairly well defined continuum
(unlike the cross-correlations produced by ``fxcor"). The width of the
broadened profile can be written as,

\begin{equation}
(\Delta\lambda)^{2} = (\Delta\lambda_{0})^{2} + (k v \sin i)^{2},
\end{equation}

\noindent
where  $\Delta\lambda$ is the fitted Gaussian FWHM of the broadened profile, 
$\Delta\lambda_{0}$ is the fitted Gaussian FWHM of the non-rotating (intrinsic) profile, and
$k$ is a constant related to units of wavelength per unit of rotation speed,
to be determined from our calibration. The ``intrinsic" profile includes the
effect of instrumental broadening, and in our case a width of 1.25 \AA~ appears
to be appropriate. This is somewhat larger than the $\sim$1.1 \AA~ reported by
\citet{sb97}, from higher resolution spectra. The calibration we
have obtained applies to our particular resolution, the Gaussian fits  and
our particular ``fxcor" setup and templates.  The constant $k$ is $0.0315$ \AA~
 km s$^{-1}$, for zero limb-darkening.

We have chosen to use zero limb darkening for the H$\alpha$ emission
rotation calibration on the assumption that chromospheric emission is
largely optically thin, or comes from a complex network of optically thick
filaments. On the other hand, the continuum around H$\alpha$ in M dwarfs is
assumed to have a limb darkening of approximately 0.8 (Gray 1992, Fig.17.6).
To study the effects of rotation on the cross-correlation measurements, we
use the spectra convolved with the rotation profile with a limb darkening of
0.8. Of course, an accurate study of rotation effects at high
signal-to-noise ratios and high resolution would require full intensity
models through lines and bands, but that is far beyond what we are doing
here, which is an initial reconaissance at moderate resolution and rather
low signal-to-noise ratio. 

The broadened spectra were measured using ``fxcor" with exactly the same
templates and filtering parameters as used for the rapidly rotating stars.
While FWHM values of the Gaussians fitted to the cross-correlation profiles
are less well defined than the H-alpha profiles, the surprising result is
that the radial velocities obtained by "fxcor" depend on the rotation
velocity, for $v \sin i > 50 $ km s$^{-1}$. 
This bias
appears clearly as a difference between the H$\alpha$ and cross-correlation
velocities for the ultrafast rotators. It is probably caused by the
asymmetric line and bandhead structure of the spectrum redward of H-alpha,
when the extra smoothing parameters are applied to these spectra. The radial
velocities of the fast rotators are corrected accordingly in column 8 of
Table~\ref{tab2} . There was no evidence for bias of radial velocity with
rotation for the slower rotators. Note that in our cross-correlation
calculations only the object stars were filtered, not the templates. The
effect of $v \sin i$ on cross-correlation radial velocities in Pleiades fast
rotators was indicated by \citet{ts00}, but not quantified.

Our technique for estimating rotation speed from H$\alpha$ appears to be
consistent for $\Delta\lambda \ge 1.4$ \AA~ and $v \sin i > 20 $ km s$^{-1}$,
but is not accurate below that, a limitation imposed by the resolution of
our spectrograph. As we shall see, that does not prevent us from some finding
interesting results. The range of $\Delta\lambda_{0} = 1.25 \pm 0.05$ \AA~
results in an uncertainty of at most $\pm 4 $ km s$^{-1}$ at $v
\sin i = 20 $ km s$^{-1}$, and less at higher rotation rates. We choose 
to leave rotation rates below $20 $ km s$^{-1}$ undefined. 

\subsection{Lithium doublet measurements}

The spectral range observed includes the Li $\lambda$6708 feature. Spectra
of each star were corrected for heliocentric velocity and averaged. The
region $\lambda\lambda$6700-6720 was plotted and examined for the lithium
feature, the equivalent width being measured when the feature appeared to be
present. The signal-to-noise ratios of the averaged spectra span a wide
range, since the apparent magnitudes of the program stars were roughly in
the range V = 12-16 and the number of observations of each star also
differed widely. In several cases, upper limits as small as 0.02 \AA~ for the
Li $\lambda$6708 equivalent width can be confidently established, in others
little can be said. Significant results will be described in the notes on
individual stars.

 
 

\begin{deluxetable}{llrrrcrrrcccrrc}
\label{tab3}
\rotate
\tabletypesize{\scriptsize}
\pagestyle{empty}
\setlength{\tabcolsep}{0.03in}
\tablecolumns{16}
\tablewidth{0pc}
\tablenum{3}
\tablecaption{Spectrophotometric Data of Active M Dwarf
Sample\label{tab3}}
\tablehead{
\colhead{Object} &                              	
\colhead{Other Names} &                            	
\colhead{V} &                            		
\colhead{$\sigma_{V}$} &                    		
\colhead{V-I} &                            		
\colhead{$\sigma_{(V-I)}$} &                    	
\colhead{M$_{V}$} &            				
\colhead{Dist.} &                       		
\colhead{BC$_{I}$} &                       		
\colhead{V-R} &                       			
\colhead{M$_{R}$} &              			
\colhead{f$_{H\alpha}$}  &                      	
\colhead{M$_{bol}$} &               			
\colhead{L$_{H\alpha}$} &                       	
\colhead{log(L$_{H\alpha}/$L$_{bol})$} \\              	
\colhead{$(RASS~Desig.)$} &                               
\colhead{$(from~SIMBAD)$} &                               
\colhead{$(mag)$} &                               	
\colhead{$(mag)$} &                       		
\colhead{$(mag)$} &                               	
\colhead{$(mag)$} &                               	
\colhead{$(mag)$} &                               	
\colhead{$(pc)$} &                               		
\colhead{$(mag)$} &                                    	
\colhead{$(mag)$} &                              		
\colhead{$(mag)$} &                              		
\colhead{($erg~s^{-1} cm^{-2}$)} &                      
\colhead{$(mag)$} &                              		
\colhead{($erg~s^{-1}$)} &                       	
\colhead{$(dex)$} \\                       		
\colhead{(1)} &
\colhead{(2)} &
\colhead{(3)} &
\colhead{(4)} &
\colhead{(5)} &
\colhead{(6)} &
\colhead{(7)} &
\colhead{(8)} &
\colhead{(9)} &
\colhead{(10)} &
\colhead{(11)} &
\colhead{(12)} &
\colhead{(13)} &
\colhead{(14)} &
\colhead{(15)} 
}
 
\startdata
RX~J0016.9+2003 &  G 131-47, GJ 3022              &  13.821 & 0.012 & 2.746  & 0.016 & 12.06 &  22.5 & 0.430 & 1.192 & 10.868 &   2.58e-13 &  9.741  &   3.09e+27  &    -4.09 \\    
RX~J0019.7+1951 &                                 &  15.460 & 0.097 & 3.144  & 0.029 & 13.39 &  25.9 & 0.296 & 1.363 & 12.027 &   2.02e-13 & 10.539  &   2.42e+27  &    -3.88 \\    
RX~J0024.5+3002 &  G 130-68, GJ 3033, LHS 1068    &  14.329 & 0.027 & 3.090  & 0.028 & 13.21 &  17.6 & 0.317 & 1.337 & 11.873 &   2.62e-13 & 10.433  &   3.13e+27  &    -3.81 \\    
RX~J0048.9+4435 &  LP 193-564                     &  13.088 & 0.011 & 2.627  & 0.013 & 11.66 &  19.3 & 0.461 & 1.148 & 10.512 &   1.17e-12 &  9.495  &   1.40e+28  &    -3.54 \\    
RX~J0050.5+2449 &  GJ 3061B, LP 350-19           &  12.368 & 0.235 & 2.748  & 0.062 & 12.07 &  11.5 & 0.430 & 1.193 & 10.877 &   4.81e-13 &  9.746  &   5.75e+27  &    -3.82 \\    
RX~J0102.4+4101 &  LP 194-16                      &  14.546 & 0.028 & 3.041  & 0.006 & 13.05 &  20.9 & 0.335 & 1.315 & 11.735 &   9.29e-14 & 10.336  &   1.11e+27  &    -4.30 \\    
RX~J0111.4+1526 &  GJ 3076, LP 467-16             &  14.238 & 0.026 & 3.480  & 0.020 & 14.51 &   9.6 & 0.144 & 1.534 & 12.976 &   1.04e-13 & 11.181  &   1.24e+27  &    -3.91 \\    
RX~J0122.1+2209 &  G 34-23, LTT 10491             &  13.012 & 0.005 & 2.998  & 0.005 & 12.90 &  10.5 & 0.351 & 1.296 & 11.604 &   2.15e-13 & 10.251  &   2.57e+27  &    -3.97 \\    
RX~J0123.4+1638 &                                 &  14.354 & 0.044 & 2.894  & 0.046 & 12.56 &  22.9 & 0.386 & 1.251 & 11.309 &   3.78e-13 & 10.043  &   4.52e+27  &    -3.81 \\    
RX~J0143.1+2101 &                                 &  13.610 & 0.018 & 2.830  & 0.009 & 12.34 &  17.9 & 0.406 & 1.225 & 11.115 &   2.63e-13 &  9.913  &   3.15e+27  &    -4.02 \\    
RX~J0143.6+1915 &                                 &  14.393 & 0.017 & 2.938  & 0.014 & 12.70 &  21.8 & 0.371 & 1.270 & 11.430 &   5.41e-13 & 10.131  &   6.47e+27  &    -3.62 \\    
RX~J0212.9+0000 &  G 159-46, GJ 3142, LHS 1358    &  13.547 & 0.019 & 2.929  & 0.010 & 12.67 &  15.0 & 0.374 & 1.266 & 11.404 &   1.39e-13 & 10.113  &   1.66e+27  &    -4.22 \\    
RX~J0219.0+2352 &  G 36-11, GJ 3150, LTT 10787    &  14.074 & 0.024 & 2.811  & 0.027 & 12.28 &  22.9 & 0.412 & 1.218 & 11.062 &   7.59e-13 &  9.874  &   9.08e+27  &    -3.57 \\    
RX~J0249.9+3345A&                                 &  13.973 & 0.009 & 2.784  & 0.012 & 12.19 &  22.7 & 0.419 & 1.207 & 10.983 &   7.98e-13 &  9.819  &   9.55e+27  &    -3.57 \\    
RX~J0249.9+3345B&                                 &  13.973 & 0.009 & 2.784  & 0.012 & 12.19 &  22.7 & 0.419 & 1.207 & 10.983 &   7.15e-13 &  9.819  &   8.55e+27  &    -3.62 \\    
RX~J0324.1+2347A&  L 1307 -15, [LH98] 62          &  10.447 & 0.001 & 2.110  & 0.002 &  9.94 &  12.6 & 0.543 & 0.981 &  8.959 & \nodata    &  8.385  &  \nodata    &  \nodata \\    
RX~J0324.1+2347B&  L 1307 -15, [LH98] 62          &  10.447 & 0.001 & 2.110  & 0.002 &  9.94 &  12.6 & 0.543 & 0.981 &  8.959 &   2.09e-12 &  8.385  &   2.51e+28  &    -3.73 \\    
RX~J0332.6+2843 &                                 &  13.835 & 0.009 & 2.927  & 0.013 & 12.67 &  17.1 & 0.375 & 1.265 & 11.405 &   3.65e-13 & 10.109  &   4.37e+27  &    -3.80 \\    
RX~J0339.4+2457 &  GJ 3241, LTT 11203,Wolf 1246   &  12.837 & 0.005 & 2.613  & 0.005 & 11.62 &  17.5 & 0.464 & 1.143 & 10.477 &   3.55e-13 &  9.466  &   4.25e+27  &    -4.07 \\    
RX~J0349.7+2419 &                                 &  14.361 & 0.013 & 2.944  & 0.005 & 12.72 &  21.3 & 0.369 & 1.273 & 11.447 &   3.45e-13 & 10.143  &   4.13e+27  &    -3.81 \\    
RX~J0442.5+2027A,B &  LP 415-345,                 &  14.101\tablenotemark{a} & 0.010 & 2.554 & 0.022 & 11.42 & 34.4\tablenotemark{a} & 0.477 & 1.122 & 10.298 &   6.63e-13\tablenotemark{a} &  9.343  &   7.93e+27  &    -3.84 \\    
RX~J0446.1+0644 &                                 &  15.164 & 0.019 & 3.093  & 0.020 & 13.22 &  24.5 & 0.316 & 1.339 & 11.881 &   1.70e-13 & 10.439  &   2.04e+27  &    -4.00 \\    
RX~J0448.7+1003 &                                 &  11.943 & 0.005 & 2.581  & 0.006 & 11.51 &  12.2 & 0.472 & 1.131 & 10.379 &   1.62e-13 &  9.399  &   1.94e+27  &    -4.43 \\    
RX~J0747.2+2957 &                                 &  12.794 & 0.009 & 2.379  & 0.011 & 10.84 &  24.6 & 0.511 & 1.064 &  9.776 &   9.59e-13 &  8.971  &   1.15e+28  &    -3.83 \\    
RX~J1002.8+4827 &  G 195-55, LHS 6180             &  15.422 & 0.028 & 3.502  & 0.029 & 14.59 &  14.7 & 0.133 & 1.546 & 13.044 &   1.03e-13 & 11.222  &   1.23e+27  &    -3.90 \\    
RX~J1038.4+4831 &  GJ 3613, LP 167-71,[GKL99] 221 &  13.454 & 0.017 & 2.735  & 0.018 & 12.02 &  19.3 & 0.433 & 1.188 & 10.832 &   5.23e-13 &  9.719  &   6.25e+27  &    -3.80 \\    
RX~J1132.7$-$2651A&                               &  12.125 & 0.005 & 2.485  & 0.008 & 11.19 &  15.4 & 0.492 & 1.098 & 10.092 &   1.86e-12 &  9.197  &   2.22e+28  &    -3.46 \\    
RX~J1132.7$-$2651B&                               &  12.125 & 0.005 & 2.485  & 0.008 & 11.19 &  15.4 & 0.492 & 1.098 & 10.092 &   2.28e-12 &  9.197  &   2.73e+28  &    -3.37 \\    
RX~J1221.4+3038A&  G 148-47, LP 320-416, Sand 57  &  14.709 & 0.018 & 3.465  & 0.007 & 14.46 &  11.2 & 0.151 & 1.526 & 12.934 &   5.08e-14 & 11.153  &   6.07e+26  &    -4.24 \\    
RX~J1221.4+3038B&  G 148-47, LP 320-416, Sand 57  &  14.709 & 0.018 & 3.465  & 0.007 & 14.46 &  11.2 & 0.151 & 1.526 & 12.934 &   6.00e-14 & 11.153  &   7.18e+26  &    -4.16 \\    
RX~J1310.1+4745 &  G 177-25, LHS 2686             &  14.626 & 0.017 & 3.222  & 0.007 & 13.65 &  15.7 & 0.264 & 1.400 & 12.250 &   9.83e-14 & 10.691  &   1.18e+27  &    -4.13 \\    
RX~J1332.6+3059 &                                 &  14.449 & 0.014 & 3.029  & 0.005 & 13.01 &  19.4 & 0.340 & 1.310 & 11.700 &   2.16e-13 & 10.313  &   2.58e+27  &    -3.94 \\    
RX~J1348.7+0406 &  Wolf 1494                      &  14.417 & 0.044 & 3.057  & 0.037 & 13.10 &  18.3 & 0.329 & 1.322 & 11.778 &   1.65e-13 & 10.368  &   1.98e+27  &    -4.04 \\    
RX~J1351.8+1247 &                                 &  12.216 & 0.007 & 2.192  & 0.008 & 10.21 &  25.2 & 0.536 & 1.005 &  9.205 &   1.05e-12 &  8.565  &   1.26e+28  &    -3.96 \\    
RX~J1359.0$-$0109 &                               &  13.612 & 0.015 & 2.641  & 0.009 & 11.71 &  24.0 & 0.457 & 1.153 & 10.557 &   5.63e-13 &  9.524  &   6.74e+27  &    -3.84 \\    
RX~J1410.9+0751 &                                 &  12.761 & 0.007 & 2.472  & 0.006 & 11.15 &  21.0 & 0.494 & 1.094 & 10.056 &   7.85e-13 &  9.169  &   9.39e+27  &    -3.84 \\    
RX~J1420.0+3902 &  GJ 3842, [GKL99] 298           &  12.383 & 0.006 & 2.367  & 0.007 & 10.80 &  20.8 & 0.513 & 1.060 &  9.740 &   1.46e-12 &  8.945  &   1.74e+28  &    -3.66 \\    
RX~J1432.1+1600 &                                 &  13.475 & 0.013 & 2.820  & 0.011 & 12.31 &  17.1 & 0.409 & 1.221 & 11.089 &   4.04e-13 &  9.893  &   4.84e+27  &    -3.84 \\    
RX~J1438.7-0257 &                                 &  13.809 & 0.012 & 2.729  & 0.016 & 12.00 &  23.0 & 0.435 & 1.186 & 10.814 &   3.69e-13 &  9.707  &   4.41e+27  &    -3.95 \\    
RX~J1447.2+5701 &                                 &  13.963 & 0.020 & 2.811  & 0.020 & 12.28 &  21.7 & 0.412 & 1.218 & 11.062 &   5.00e-13 &  9.874  &   5.99e+27  &    -3.75 \\    
RX~J1459.4+3618 &                                 &  15.039 & 0.041 & 3.126  & 0.043 & 13.33 &  22.0 & 0.303 & 1.354 & 11.976 &   2.57e-13 & 10.504  &   3.07e+27  &    -3.79 \\    
RX~J1509.0+5904 &  StM 219                        &  12.769 & 0.008 & 2.461  & 0.007 & 11.11 &  21.5 & 0.496 & 1.090 & 10.020 &   7.44e-13 &  9.146  &   8.90e+27  &    -3.87 \\    
RX~J1512.6+4543 &  G 179-20, GJ 3898, LHS 3035    &  13.047 & 0.010 & 2.877  & 0.008 & 12.50 &  12.9 & 0.391 & 1.244 & 11.256 &   2.89e-13 & 10.008  &   3.46e+27  &    -3.94 \\    
RX~J1523.8+5827 &  G 224-65                       &  14.310 & 0.017 & 2.864  & 0.011 & 12.46 &  23.5 & 0.395 & 1.239 & 11.221 &   4.40e-13 &  9.982  &   5.26e+27  &    -3.77 \\    
RX~J1529.0+4646 &                                 &  14.862 & 0.027 & 3.120  & 0.028 & 13.31 &  20.4 & 0.305 & 1.351 & 11.959 &   1.21e-13 & 10.492  &   1.45e+27  &    -4.12 \\    
RX~J1542.3+5936 &                                 &  15.368 & 0.044 & 3.179  & 0.040 & 13.51 &  23.6 & 0.281 & 1.379 & 12.131 &   2.42e-13 & 10.607  &   2.89e+27  &    -3.78 \\    
RX~J1547.4+4507A&  G 179-55, LP 177-102           &  13.988\tablenotemark{a} & 0.009 & 2.915  & 0.010 & 12.63 &  18.7 & 0.379 & 1.260 & 11.370 &   1.62e-13\tablenotemark{a} & 10.085  &   1.94e+27  &    -4.16 \\    
RX~J1547.4+4507B&  G 179-55, LP 177-102           &  13.988\tablenotemark{a} & 0.009 & 2.915  & 0.010 & 12.63 &  18.7 & 0.379 & 1.260 & 11.370 &   2.46e-13\tablenotemark{a} & 10.085  &   2.94e+27  &    -3.98 \\    
RX~J1548.0+0421 &                                 &  13.746 & 0.011 & 2.656  & 0.006 & 11.76 &  24.9 & 0.454 & 1.158 & 10.602 &   4.09e-13 &  9.556  &   4.89e+27  &    -3.97 \\    
RX~J1648.0+4522 &                                 &  13.669 & 0.008 & 2.877  & 0.026 & 12.50 &  17.1 & 0.391 & 1.244 & 11.256 &   3.90e-13 & 10.008  &   4.66e+27  &    -3.81 \\    
RX~J2137.6+0137 &  2E 4498, EUVE J2137+01.6       &  13.362 & 0.006 & 2.832  & 0.006 & 12.35 &  15.9 & 0.405 & 1.226 & 11.124 &   7.50e-13 &  9.917  &   8.97e+27  &    -3.56 \\    
RX~J2227.8$-$0113 &                               &  13.432 & 0.008 & 2.629  & 0.010 & 11.67 &  22.5 & 0.460 & 1.149 & 10.521 &   9.80e-13 &  9.500  &   1.17e+28  &    -3.61 \\    
RX~J2243.7+1916 &                                 &  13.083 & 0.007 & 2.488  & 0.005 & 11.20 &  23.8 & 0.491 & 1.099 & 10.101 &   2.30e-13 &  9.203  &   2.75e+27  &    -4.36 \\    
RX~J2317.5+3700 &                                 &  11.690 & 0.002 & 2.048  & 0.001 &  9.73 &  24.7 & 0.547 & 0.962 &  8.768 &   4.28e-13 &  8.247  &   5.13e+27  &    -4.47 \\    
RX~J2326.2+2752 &                                 &  12.219 & 0.004 & 2.448  & 0.004 & 11.07 &  17.0 & 0.499 & 1.086 &  9.984 &   1.21e-12 &  9.118  &   1.45e+28  &    -3.67 \\    
RX~J2333.3+2714 &  G 128-76, LTT 16928            &  13.303 & 0.007 & 2.566  & 0.003 & 11.46 &  23.4 & 0.475 & 1.126 & 10.334 &   2.87e-13 &  9.368  &   3.43e+27  &    -4.20 \\    
RX~J2337.5+1622 &                                 &  16.144 & 0.046 & 3.687  & 0.045 & 15.20 &  15.4 & 0.033 & 1.650 & 13.550 &   4.10e-14 & 11.563  &   4.91e+26  &    -4.16 \\    
RX~J2349.2+1005 &  G 30-18, GJ 4363, LTT 17016    &  13.567 & 0.006 & 2.647  & 0.007 & 11.73 &  23.3 & 0.456 & 1.155 & 10.575 &   2.43e-13 &  9.537  &   2.91e+27  &    -4.20 \\    
RX~J2354.8+3831 &                                 &  13.187 & 0.007 & 2.791  & 0.009 & 12.21 &  15.7 & 0.417 & 1.210 & 11.000 &   4.29e-13 &  9.834  &   5.14e+27  &    -3.84 

\tablenotetext{a}{Corrected for binarity}
\tablecomments{The V,I photometry and photometric distances are from \citet{f98}}   
\enddata
\end{deluxetable}

\section{RESULTS}
\label{sec3}

The stars in our sample, as summarised in Table~\ref{tab2} and
Table~\ref{tab3}, fall into the following categories:

\subsection{Slowly rotating single stars}

The ``typical" object in our sample is a dMe with a relatively narrow 
H$\alpha$ emission line
( with $\Delta\lambda \lesssim 1.4$\AA~  corresponding to $v
\sin i \lesssim 20 $ km s$^{-1}$ )  of equivalent width $\sim 5$ \AA. Since
we cannot determine $v \sin i$ much less than $20 $ km s$^{-1}$, we
 place our threshold at this value. The standard deviation of the FWHM
of the Gaussians fitted to the H$\alpha$ profiles should indicate the
reliability of the $v \sin i$ determination, since we have already seen that
the calibration is reasonably consistent. 

\subsection{Fast and ultra fast rotators}

\begin{figure}
\figurenum{3}
\plotone{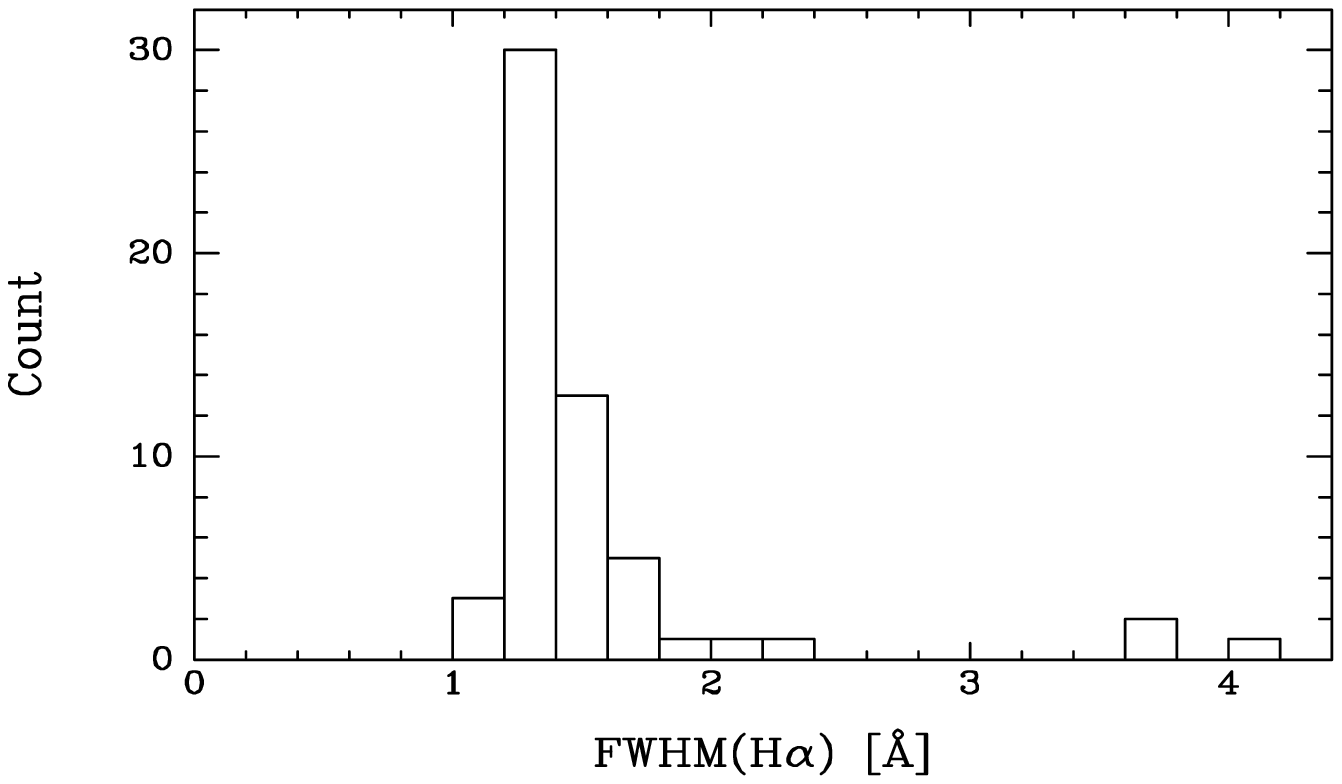}
\figcaption[fig3.eps]{\label{fig3}
Distribution of H$\alpha$ emission profile full width at half maximum. This
shows the population of fast rotators ( FWHM $> 1.6$ \AA, i.e. $v \sin i
> 30 $ km s$^{-1}$) and ultra-fast rotators.}
\end{figure}

While there is a considerable number of stars with apparent $v \sin i$ in
the range 30-60 km s$^{-1}$, which we designate as Fast Rotators, there is a
striking group of three Ultra Fast Rotators: RX~J0143.6+1915, RX
J1420.0+3902 and RX~J2227.8-0113, with estimated $v \sin i$ values of 121,
109 and 112  km s$^{-1}$ respectively. Figure~\ref{fig3} shows the distribution 
of the FWHM of H$\alpha$, and hence $v \sin i$,  in the Fleming sample. 

\begin{figure}
\figurenum{4}
\plotone{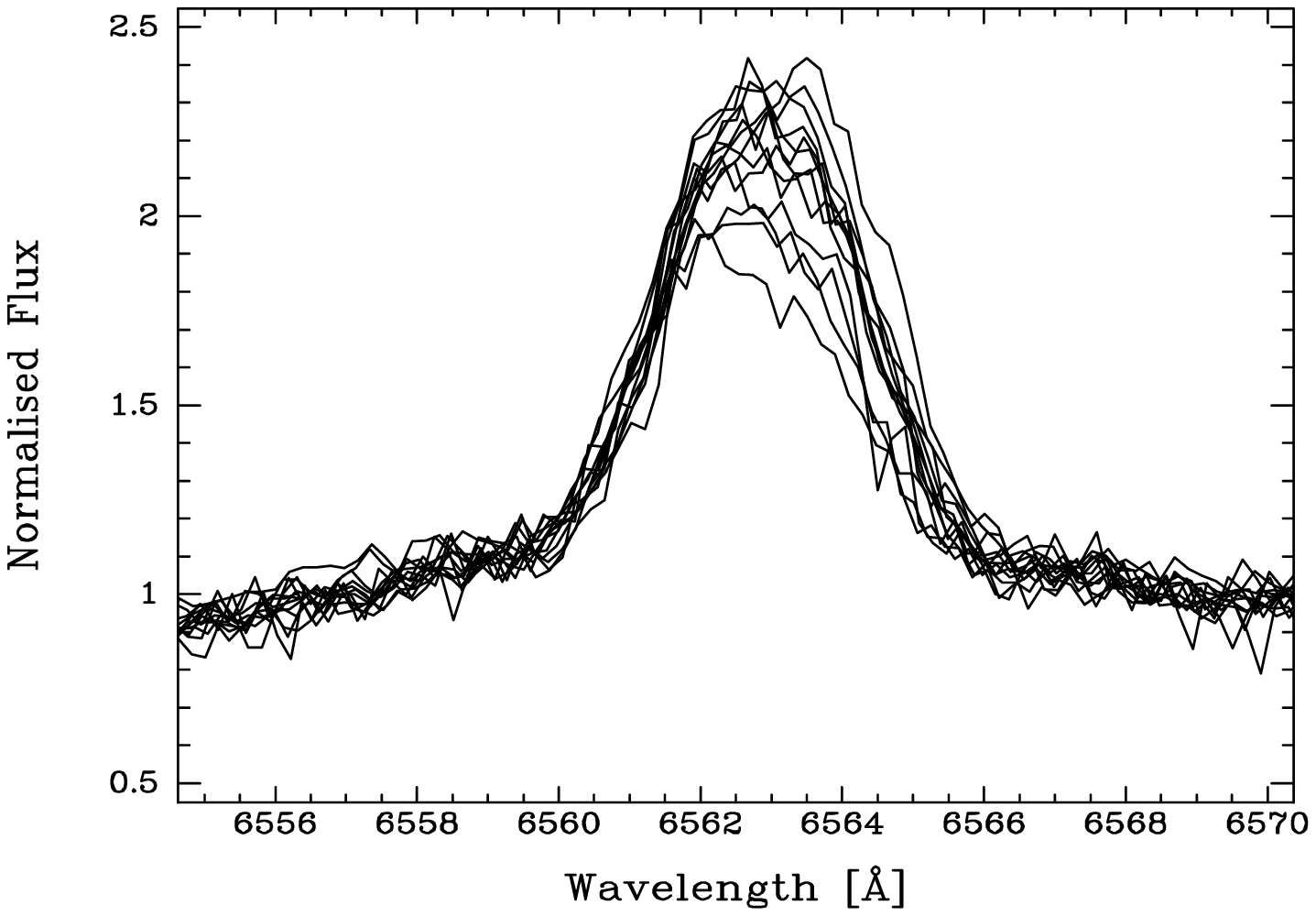}
\figcaption[fig4.eps]{\label{fig4}
Variation of the H$\alpha$ profile of  the ultra-fast rotator RX
J1420.0+3902. The plots have been scaled to the same continuum level, with
zero flux at the bottom of the plot. The S/N of each spectrum is different
due to different total integrations of flux.}
\end{figure}

\begin{figure}
\figurenum{5}
\plotone{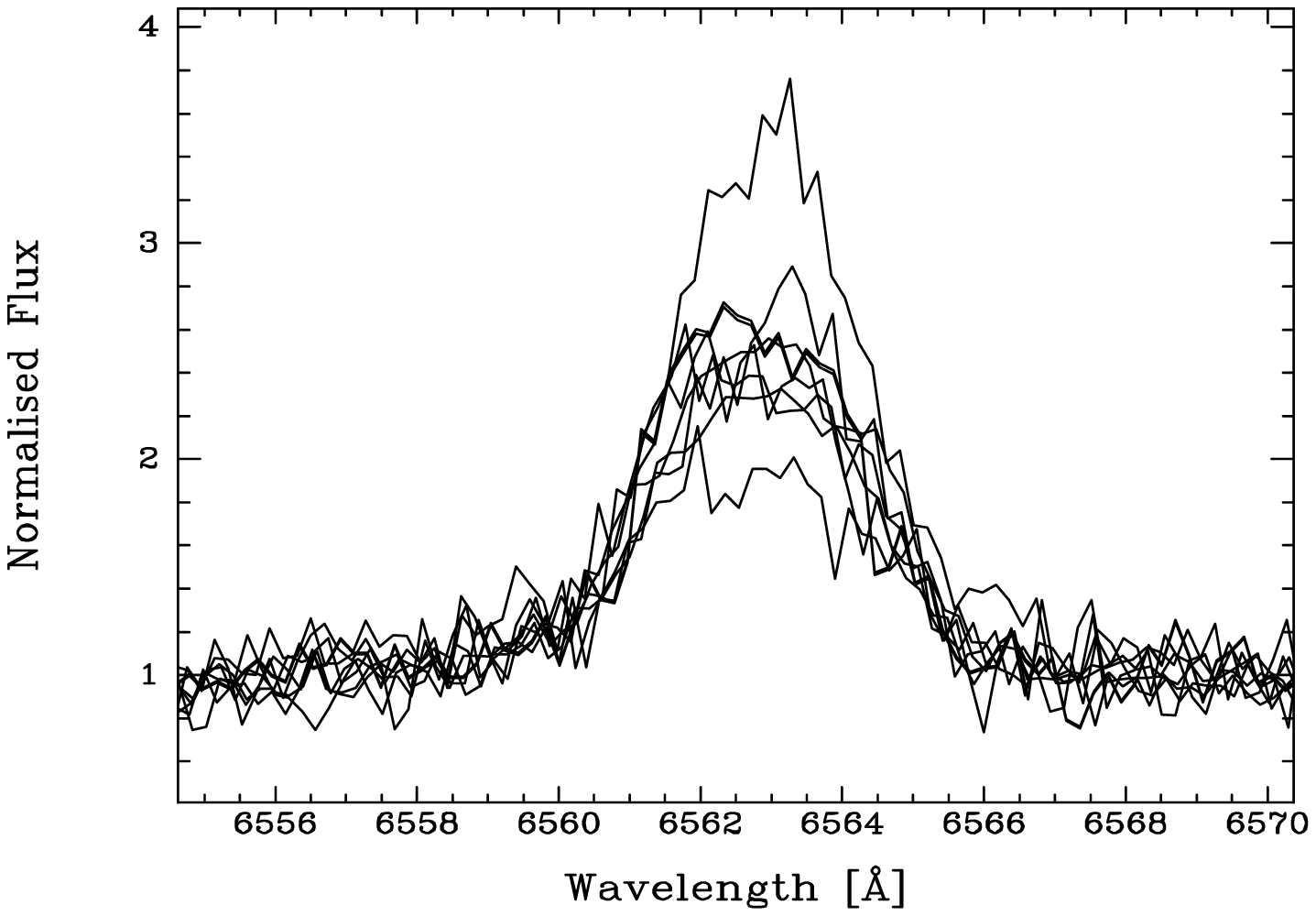}
\figcaption[fig5.eps]{\label{fig5}
Variation of the H$\alpha$ profile of  the ultra-fast rotator 
RX~J2227.8-0113. Same description as Figure~\ref{fig4}. }
\end{figure}

The radial velocities of the UFRs are not as easy to determine, as noted
above, but there is no evidence that these objects are either spectroscopic
binaries or doubles (from viewing DSS and 2MASS images), although 2MASS
imagery of RX~J2227.8-0113 shows some faint companion objects within
15\arcsec~ or so. The UFRs are intrinsically variable, as seen in the
H$\alpha$ line in Figures~\ref{fig4} and \ref{fig5}. Both H$\alpha$ emission
profiles and absorption line spectra are consistent with rotation as the
cause of the observed line broadening.

\begin{figure}
\figurenum{6}
\plotone{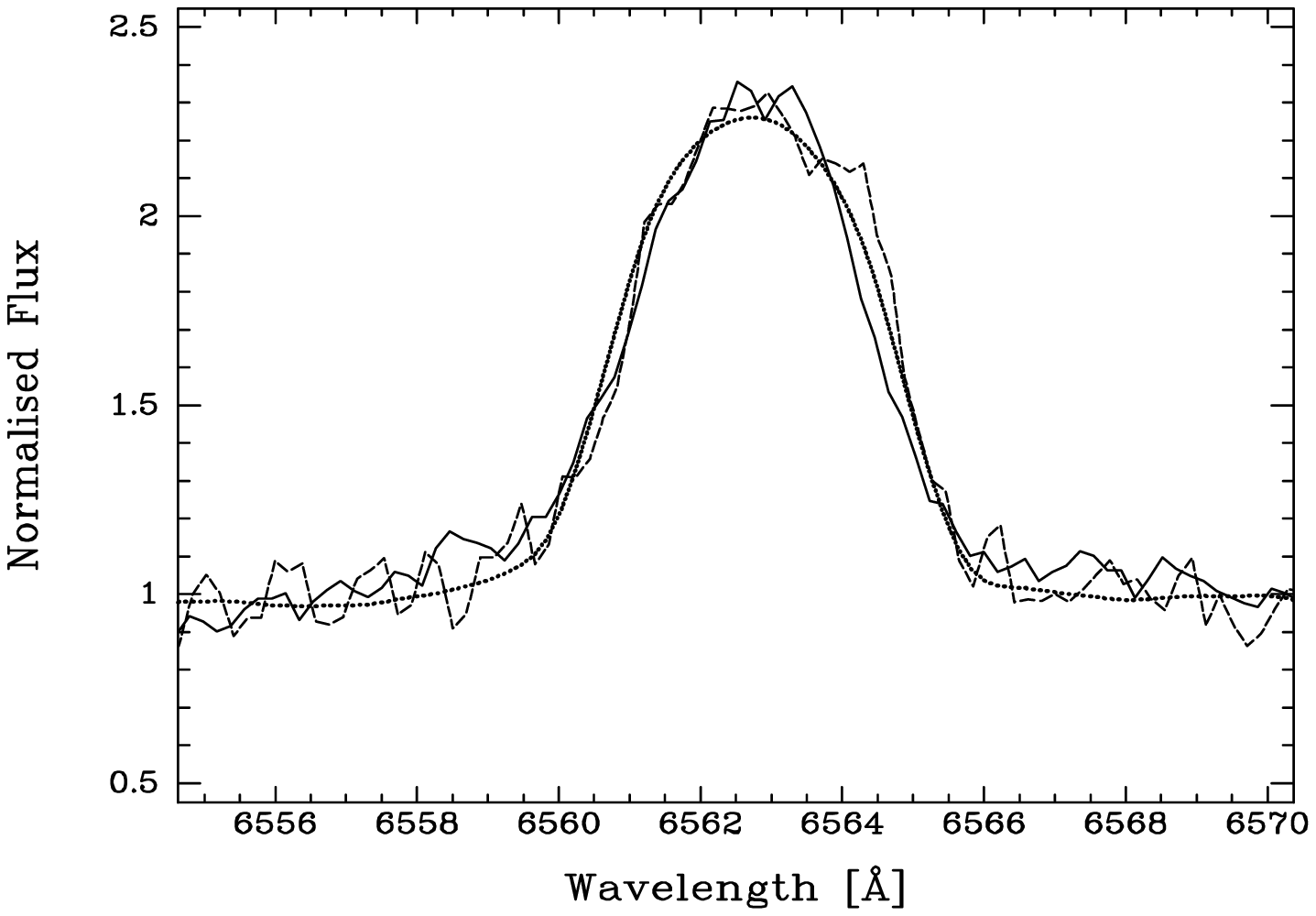}
\figcaption[fig6.eps]{\label{fig6}
Ultra fast rotators and modelled H$\alpha$ profile. The solid line
represents the average spectrum of RX J1420.0+3902 and the dashed line is RX
J2227.8-0113, scaled to their continua. The dotted line represents the
spectrum of RX J1038.4+4831 artificially broadened to $v \sin i = 110$ km
s$^{-1}$, with emission portion above continuum scaled up (see text).
}
\end{figure}

The results of our rotation analysis for two UFRs are shown in
Figure~\ref{fig6}. The mean spectrum of R XJ1038.4+4831 was broadened to
$110$ km s$^{-1}$ using equation (1). The part of the profile above a flat
level joining the continuum at $\lambda\lambda6557.66-6568.06$\AA~  was scaled
up by a factor of 1.36 to better fit the average emission strength of the
UFRs RX J1420.0+3902 and RX J2227.8-0113. This broadened spectrum with
scaled emission is drawn as a dotted line. The mean spectra of the UFRs were
shifted to remove the predicted bias of $+8$ km s$^{-1}$. The uncertainty of
our $v \sin i$ estimates for the UFRs appears to be about $\pm 10$ km s$^{-1}$.

\subsection{Spectroscopic binaries}

\begin{figure}
\figurenum{7}
\plotone{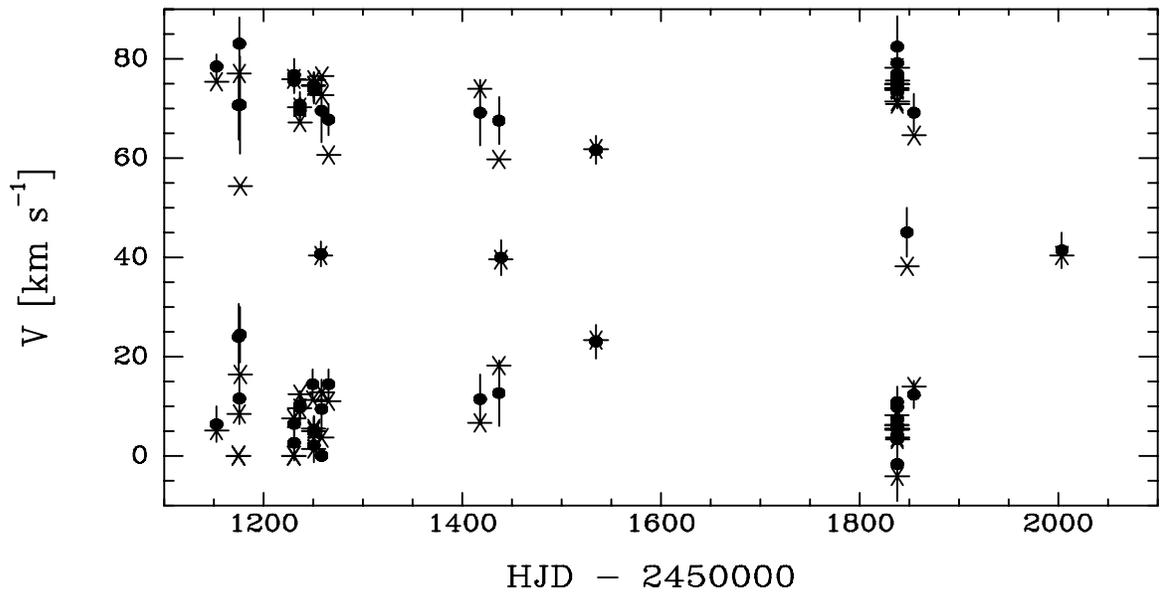}
\figcaption[fig7.eps]{\label{fig7}
Radial velocities of the double-lined double-dMe spectroscopic binary
RX~J0442.5+2027. Solid dots are cross-correlation measurements, asterisks
are H$\alpha$ Gaussian fits. A reliable period has not yet been determined.}
\end{figure}

Two stars were quickly found to be double-lined double-emission
spectroscopic binaries, with periods of a few days: RX~J0442.5+2027 (LP
415-345) and RX~J1547.4+4507 (G179-55, LP 177-102). The first was noted as an
SB2 and Hyades member by \citet{sb97}, and therefore was not as extensively
observed by us as the second. This is unfortunate, because it appears that
its period is difficult to determine (Figure~\ref{fig7}). Radial velocities
from H$\alpha$ and cross-correlation are presented in Table 1. A possible
period is around 5.595 days, with values of half or a quarter of that being
possible. More observations are required.

The second star, RX~J1547.4+4507, does not appear to have been previously
known as an SB2. Velocities are given in Table 1. Application of the IRAF
task ``pdm" and a version of the spectroscopic binary program SBBG
\citep{bg69}  yields the elements in Table~\ref{tab4} . We find that the
orbit is circular with a period of 3.55 days. 

The objects RX~J0102.4+4101, RX~J1332.6+3059 and RX~J2317.5+3700 have
possible radial velocity variability but more observations are needed to
establish their nature.  RX~J2349.2+1005 appears to show long-term radial
velocity variability.

 
 

\begin{deluxetable}{lcccc}
\label{tab4}
\pagestyle{empty}
\tabletypesize{\scriptsize}
\tablecolumns{5}
\tablewidth{0pt}
\tablenum{4}
\tablecaption{Orbital Elements of RX~J1547.4+4507\label{tab4}}
\tablehead{
\colhead{Element} &	
\colhead{Symbol} &      
\colhead{Value} &	
\colhead{Std.Err.} &	
\colhead{Unit} \\	
\colhead{(1)} &
\colhead{(2)} &
\colhead{(3)} &
\colhead{(4)} &
\colhead{(5)} 
}
\startdata
Epoch of periastron & $T_{0}$ & 2451262.375 & 0.005 & HJD \\
Period & P & 3.54997 & 0.00005 & days \\
Semi-Amplitude (A) & $K_{1}$ & 56.1 & 0.4 & km s$^{-1}$ \\
Semi-Amplitude (B) & $K_{2}$ & 55.9 & 0.5 & km s$^{-1}$ \\
Systemic Velocity & $V_{0}$ & -21.8 & 0.4 & km s$^{-1}$ \\
Eccentricity & $\epsilon$ & 0.008 & 0.007 & \nodata \\
Longitude of Periastron & $\omega$ & 197.9 & 42 & degrees \\
Mass (Component A) & $M_{1} \sin^{3} i$ & 0.257 & 0.005 & M$_{\sun}$ 
\enddata
\end{deluxetable}
 

\subsection{Visual doubles}

A number of stars in this sample are either known doubles, or we have
observed them to be doubles. A perusal of 2MASS and DSS images, combined
with our observations and some other published or archival data, yields the
following doubles: RX~J0050.5+2449 (unresolved in our spectra),
RX~J0123.4+1638 (only component A observed), RX~J0249.9+3345,
RX~J0324.1+2347, RX~J0349.7+2419 (only A observed), RX~J1002.8+4827 (only A
observed), RX~J1132.7-2651, RX~J1221.4+3038, RX~J1509.0+5904 (only A
observed).  Thanks to the improvement of seeing at the David Dunlap
Observatory in recent years, we were able to resolve several of these
doubles in the course of our spectroscopy. It means that the photometric
distances derived by \citet{f98} for these objects are uncertain due to the
contribution of the companion stars to their observed magnitudes and
colours. CCD observations at high angular resolution are needed to derive
reliable magnitudes and colours for these stars.

\section{RESULTS FOR INDIVIDUAL STARS}

\subsection{RX~J0050.5+2449}

DSS and 2MASS images suggest that this object may be a close double.
Examination of archival CFHT H-band AO images taken in 2000 August by Jewitt
and Sheppard shows that this is an unequal double, with a separation
of about 1.83\arcsec. \citet{rhg95} measured a radial velocity of 9.2 km
s$^{-1}$, compared with the $6 \pm 1$ km s$^{-1}$ we obtain.

\subsection{RX~J0102.4+4101}

At first glance this looks like a rapid rotator, but more likely is also an
SB2, since its absorption cross-correlation profiles are relatively narrow
and show significant velocity variation and possible multiplicity. The
cross-correlation and H$\alpha$ velocities also strongly disagree. It has a
brighter common proper motion companion about 30\arcsec~  distant, LP 194-15
\citep{w91}. 

\subsection{RX~J0143.1+2101}

We have only one spectrum of this object, and more observations are needed.
The observed radial velocity is about $-46 $ km s$^{-1}$, the most extreme
in the Fleming sample if it is not variable. There is no fast rotation
evident.

\subsection{RX~J0143.6+1915}

This is an ultra fast rotator, with $v \sin i \simeq 120 $ km s$^{-1}$. More
observations are needed to examine its variability. 

\subsection{RX~J0249.9+3345 A \& B}

This is a close visual double in which both components are rapid rotators
with $v \sin i \sim 35-50 $ km s$^{-1}$. More observations in conditions of
good seeing are needed.

\begin{figure}
\figurenum{8}
\plotone{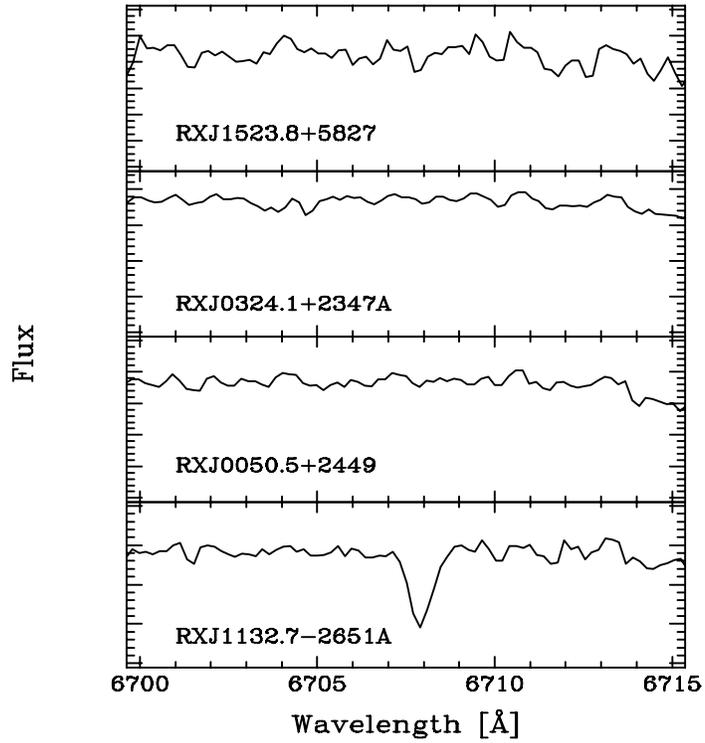}
\figcaption[fig8.eps]{\label{fig8}
Lithium line region, averaged spectra. Zero flux is at the bottom of each
plot. The only unequivocal detection is in the bottom plot (see text). The
top plot represents a possible detection, but the spectrum is too noisy for
certainty. The average spectrum of RX~J0324.1+2347A clearly does not show
the strong Li absorption reported by \citet{lh98}.}
\end{figure}

\subsection{RX~J0324.1+2347 A \& B}

Only the less luminous, cooler B component has H$\alpha$ in emission,
although there is some filling in of the H$\alpha$ absorption in component
A. Neither component shows rapid rotation. Component A appears to have Li
$\lambda 6708$ with an equivalent width of at most 0.02 \AA~
(Figure~\ref{fig8}), which does not agree with the value of 0.19 \AA~
measured by \citet{lh98}. Component B appears to have variable radial
velocity. \citet{w91} notes that this double has a common proper motion
companion, L1307-14 C, 100\arcsec~ away.

\subsection{RX~J0442.5+2027}

This double-lined, double-emission spectroscopic binary appears to be at a
distance of 34 pc, which puts it at about the tidal radius of the Hyades
cluster. In Table~\ref{tab3}, the components are assumed to be equal in
magnitude and emission strength; the magnitudes and equivalent widths have
been corrected accordingly. The H$\alpha$ emission from each component is
completely normal for  stars in this sample.

\subsection{RX~J0446.1+0644}

This star has a velocity quite similar to that of the Hyades, although it is
some distance from the cluster on the sky and is closer \citep{f98}.

\subsection{RX~J0448.7+1003}

\begin{figure}
\figurenum{9}
\plotone{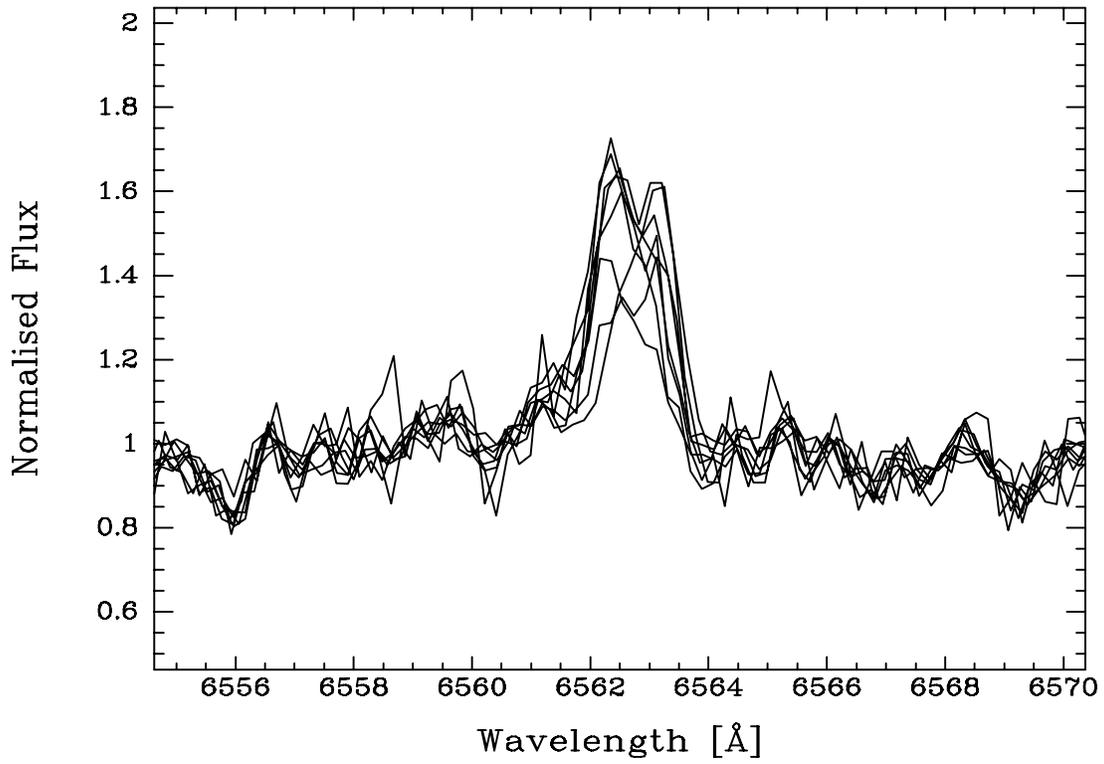}
\figcaption[fig9.eps]{\label{fig9} 
Variation of the H$\alpha$ profile of  the variable object RX~J0448.7+1003.
This is probably a spectroscopic binary. The H$\alpha$ emission is
relatively quite weak.}
\end{figure}

This has a rather weak, asymmetric and variable H$\alpha$ emission, shown in
Figure~\ref{fig9}, with significant H$\alpha$ radial velocity variation and
discrepancy between absorption cross-correlation and H$\alpha$ velocities.
It may be rotating with $v \sin i \sim 20 $ km s$^{-1}$, but our resolution is
insufficient to assure that we are seeing the effects of rotation combined
with irregular surface emission. This is a somewhat unusual star in this
sample, and probably a binary.

\subsection{RX~J1038.4+4831}

\begin{figure}
\figurenum{10}
\plotone{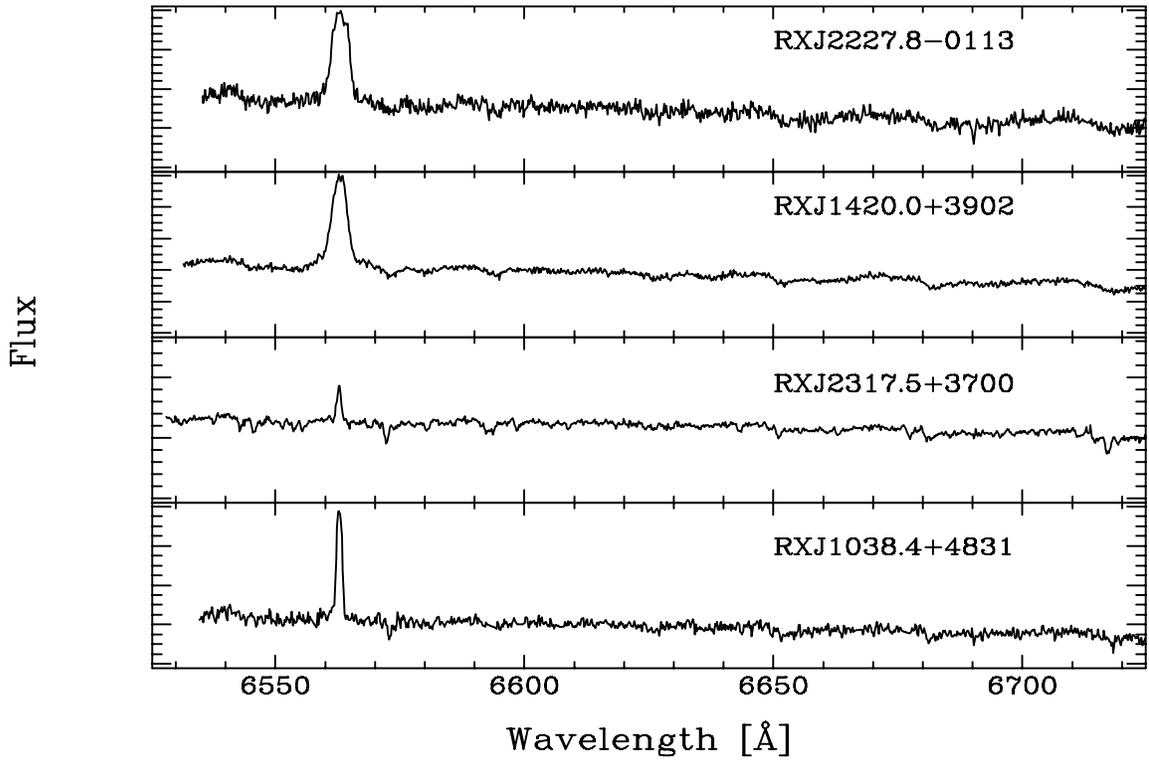}
\figcaption[fig10.eps]{\label{fig10}
Examples of averaged spectra. Zero flux is at the base of each plot. The top
two plots show ultra-fast rotators. The star RX~J2317.5+3700 has the weakest
H$\alpha$ emission in the sample, while RX~J1038.4+4831 is a well-observed
star without detected variability or fast rotation. These spectra have not
had telluric features removed.}
\end{figure}

This was the best observed apparently non-binary slowly rotating star in our
sample, and was therefore used as the template for our rotation model
calibration, and is shown in Figure~\ref{fig10}. Of course, higher resolution
observations may show that this star has significant rotation, but less than
$20 $ km s$^{-1}$.

\subsection{RX~J1132.7-2651}

This is a double or multiple system, also known as TWA 8 \citep{wz99}.
 This system is thought to be a member of the TW Hydrae association, a
group of T Tauri stars perhaps 10 Myr old and 50 pc distant \citep{wz99,
td00}. We measured the velocity of TWA 8A to be
$7.3\pm1.3 $ km s$^{-1}$, which is significantly less than the mean velocity of
$11.30 \pm 0.56 $ km s$^{-1}$ for the centres of mass of 6 well studied single or
multiple members of the TWA given by \citet{tnl01},
but in agreement with the velocity of $7.5 $ km s$^{-1}$ for TWA~8 
itself in Table 6 of
\citet{td00}. \citet{kz97} found a molecular cloud velocity of
about $12.2 $ km s$^{-1}$ in the direction of TW Hya, with a velocity
dispersion of $0.7 $ km s$^{-1}$ or so, similar to the stellar velocity
dispersion of the six systems of \citet{tnl01} and the internal dispersion of 
$0.8 $ km s$^{-1}$
proposed by \citet{mf01}. The latter authors also position
the center of mass of the TWA at a distance of 73 pc in the direction of TW
Hya, with the nearest members of the TWA as close as 17.5 pc from the Sun.
They propose a dynamical age of 8.3 Myr, similar to the age of 10 Myr or
so suggested by many other authors. They furthermore propose a linear
expansion of $0.12 $ km s$^{-1}$ pc$^{-1}$ away from the center of mass.

The fainter B component, about 13\arcsec~  to the south, was measured to
have an H$\alpha$ velocity of $6.4\pm1.4 $ km s$^{-1}$, with the absorption
cross-correlation velocity not being very reliable due to the star's
faintness. There is a third, fainter object closer to component A, but much
too faint for us to observe. \citet{lbs01} find no close brown dwarf
companion around TWA 8B. There does not appear to be a proper motion
published for TWA 8; it has been somewhat neglected in the literature
compared with its TWA fellow travellers. We measured a lithium equivalent
width in component A of $0.54 \pm 0.03$\AA, in excellent agreement with
\citet{td00} and \citet{wz99}. The width of the lithium line (see
Figure~\ref{fig8}) is about 1.5 \AA~ and its FWHM is about 0.7 \AA, which is
consistent with a rotation speed somewhat less than the $33 $ km s$^{-1}$
inferred from the H$\alpha$ width; the FWHM of H$\alpha$ for this star has a
relatively large standard deviation in our measurements.

The evidence is quite strong that RX~J1132.7-2651  is a member of the TW
Hydrae Association. The distance to this system is important. \citet{f98}
simply used the M dwarf V-I color-luminosity relation of \citet{spi89} to
estimate $M_{V} = 11.19$, deriving a distance of 15.4 pc.  Since TWA 8
consists of at least two pre-main sequence stars, its luminosity is larger
than assumed by Fleming.  \citet{wz99} report K magnitudes of 7.44 and
9.01 respectively for TWA 8 A \& B. \citet{td00} infer
$M_{V} = 8.74$ for TWA 8A, at a distance of 50 pc. The gradient of
\citet{mf01} predicts the observed radial velocity if the distance
is about 25-35 pc, corresponding to $M_{K} \simeq 4.7-5.4$ for TWA 8A. This
is somewhat less luminous than suggested by Figure 3 of \citet{wz99}.
More detailed study of this double, or possibly triple, system is needed. In
particular, its proper motion and trigonometric parallax should be measured.

\subsection{RX~J1310.1+4745}

This star (LHS 2686, G177-25) was observed by \citet{gr97}, who found a
radial velocity of $-15.8 $ km s$^{-1}$, in good agreement with our value of
about $-14 $ km s$^{-1}$.

\subsection{RX~J1420.0+3902}

This star is our best-observed ultra fast rotator, with $ v \sin i \simeq
109 $ km s$^{-1}$. Our observations show clear variability of the H$\alpha$
profile (Figure~\ref{fig4}). The radial velocity measurements show scatter,
but there is no evidence of binarity. Our mean de-biased radial velocity of
about $-21 $ km s$^{-1}$ disagrees with the velocity of $3.5 $ km s$^{-1}$
quoted by \citet{rhg95}.

\begin{figure}
\figurenum{11}
\plotone{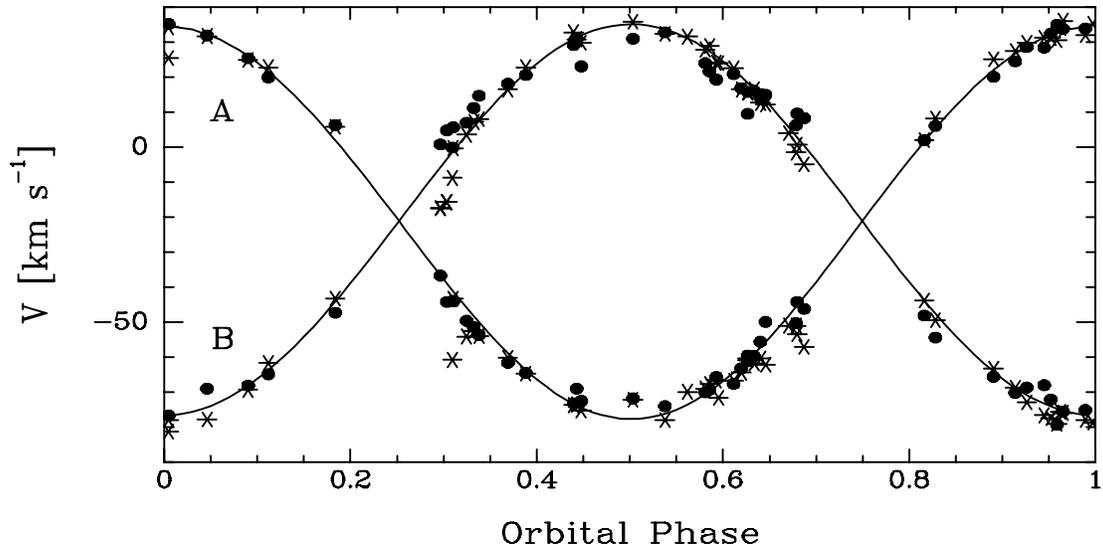}
\figcaption[fig11.eps]{\label{fig11}
Radial velocity curves for RX~J1547.4+4507. Solid dots are cross-correlation
velocity measurements, stars are H$\alpha$ measurements. Unresolved profiles
have been excluded from the fit. Components are labelled A and B as in Table
1. The orbital phase is referred to the epoch in Table 4 plus the longitude
of periastron expressed in days.}
\end{figure}

\subsection{RX~J1547.4+4507}

The fitted orbital elements are shown in Table~\ref{tab4}. Note that the
eccentricity is zero within the error range, so the orbit is circular. The
angular distance $\omega$ between the ascending node and periastron is
purely formal; the epoch of zero phase is Figure~\ref{fig11} shows the
cross-correlation and H$\alpha$ velocities for each component only when when
both components A and B are resolved. The components have equal masses.

Fleming (1998)  observed a V-I color of 2.915 for this system, and used  the
color-luminosity relation of \citet{spi89} to deduce a distance of 13.3 pc,
assuming this was a single star. If components A and B have identical
luminosities, the distance should really be $\sqrt 2$ times greater, i.e.
18.8 pc.  Using Fleming's color-index derived value of $M_{V} = 12.63$ for
each component, the mass can be inferred using the relationship of
\citet{hm93},

\begin{equation}
\log (M/M_{\sun}) =  -0.1681 M_{V} + 1.4217
\end{equation}  

\noindent
for $0.50 \ge M \ge 0.18 M_{\sun}$, giving $M = 0.20\pm0.04 M_{\sun}$ for
each component. On the other hand, the component masses for an orbital
inclination of 90\arcdeg~ are relatively large ($0.26\pm0.005 M_{\odot}$),
suggesting that the orbit may be close to edge-on, and therefore worthy of
monitoring for possible eclipses, although the probability of these
occurring is at most a few percent.  The components are definitely not
rotating rapidly, which is consistent with the orbit having been
circularized by tidal friction; photometric monitoring will show whether
surface (spot) rotation on each component is synchronous with the orbit. The
mass ratio is 1.00 within our observational error.

The components identified as A and B in Table 1 and Figure~\ref{fig11} have
unequal average H$\alpha$ equivalent widths of 1.24 and 1.89 \AA~
repectively, with standard deviations of 0.5 and 0.4 \AA, excluding
unresolved profiles. It is interesting that two M dwarfs of equal mass
in a circular orbit have unequal emission strengths. After doubling each
equivalent width to allow for the continuum dilution by the other component,
the $L_{H\alpha}/L_{bol}$ ratio for each component is normal for this
sample.

\subsection{RX~J2137.6+0137}

This is a fast rotator, with $v \sin i \simeq 55 $ km s$^{-1}$. There is some
scatter in the radial velocities, but the H$\alpha$ profile does not show
the dramatic variability of the UFRs. It possibly has been detected as an
EUVE source \citep{cc99}.  There are 3 faint stars within about 20\arcsec~  in
XDSS images.

\subsection{RX~J2227.8-0113}

We obtained 8 spectra of this UFR ($v \sin i \simeq 112 $ km s$^{-1}$).
H$\alpha$ variability is even more pronounced than in RX~J1420.0+3902, which
has a similar spectrum (Fig.~\ref{fig10}), although active flare stars
certainly should display great Balmer line variability \citep{mz80}.  There
is a possible companion about 11\arcsec~ to the West, and about 10 faint
objects within 30\arcsec, as seen in DSS and XDSS images, most also seen in
2MASS images. The continuum is too noisy in our data and the rotational
broadening is too great for evidence of lithium to be seen. There is
considerable scatter in our radial velocity measurements.

\subsection{RX~J2317.5+3700}

This star has the weakest and narrowest H$\alpha$ emission in the sample
(see Figure~\ref{fig10}), with a consistent velocity discrepancy
$V(fxcor)-V(H\alpha) \simeq -23 $ km s$^{-1}$. It has 3 faint companions
within 20 \arcsec~, which also appear in the 2MASS images but are clearly
bluer. Numerous other nearby faint objects are seen in the XDSS images. This
star is the bluest in Fleming's photometry, although it is probably cooler
than RX~J0324.1+2347A, which was not resolved from its companion by
\citet{f98}. It is unlikely that the H$\alpha$ emission has a nebular origin
because it is seen only in the stellar image and not more extended along the
slit. 

\subsection{RX~J2349.2+1005}

Although we have only four spectra of this star, and in the most recent
observation the velocity of H$\alpha$ could not be accurately determined due
to misplacement of the grating position, it is quite likely that the radial
velocity is variable, which would be caused by an unseen companion.  It is a
moderately fast rotator. It is star 461 in the flare star catalog of
\citet{gk99}.

\section{DISCUSSION AND CONCLUSIONS}
\label{sec4}

We set out to do a spectroscopic investigation of Fleming's  sample of X-ray
bright M dwarfs with apparent photometric distances of less than 25 pc
\citep{f98}. This was by far the faintest sample of stars  ever to be studied
spectroscopically with the David Dunlap Observatory's 1.88 m reflector, an
ambitious goal for an undergraduate practical astronomy class; many of the
important papers in this field in recent years have involved 10-metre class
telescopes at sites well away from and above the sort of environment we have
near Toronto.  Coverage of the sample was uneven due to distribution around
the sky and the weather. Nevertheless, our sensitivity, resolution and
instrumental stability were sufficient to draw some interesting conclusions.

All of the X-ray sources studied by \citet{f98} are dMe stars, though in
one or possibly two cases the H$\alpha$ emission comes from a fainter
companion unresolved by Fleming. Emission equivalent widths are generally
less than 10 \AA. Observations by \citet{hgr96} and more
recent work by \citet{mm01} suggest that dMe stars are more luminous
than dM stars with the same color, and therefore larger. This means
that Fleming's photometric distances are under-estimated, particularly since
some of the stars may be quite young and therefore well above the main
sequence. 

At this stage, a full proper motion study of these stars remains to be
completed, although \citet{f98} has shown some preliminary results. The
radial velocity dispersion of the sample, about $17 $ km s$^{-1}$, agrees
with the single-component fits to the H$\alpha$ emission sample of stars in
the {\it Third Catalogue of Nearby Stars} \citep{gj91}, observed by
\citet{rhg95}. Further studies of motions should identify streams of
different ages, as dramatically shown in the discovery of the TW Hya
Association \citep{kz97,wz99}.

Several of these stars show fast or ultra-fast rotation. Out
of 54 $systems$ in Fleming's sample, 3 are UFRs with $v \sin i \gtrsim 100
 $ km s$^{-1}$. About 10 more stars are fast rotators with $30 $ km s$^{-1}
\lesssim v \sin i \lesssim 60 $ km s$^{-1}$, including the only star in the
sample with strong lithium absorption. While the star with lithium
(RX~J1132.7-2651) is probably a member of the TW Hydrae association
and therefore less than 10 Myr in age, the other fast and ultra-fast
rotators are probably also quite youthful, especially when compared with the
approximately 110 Myr age of the Pleiades and its rapid rotators
\citep{ts00}. \citet{rm00} showed that $v \sin i$ of M dwarfs in the Hyades
was at most 35 km s$^{-1}$, with a maximum of around 100 km s$^{-1}$ in the
Pleiades. If M dwarf rotation slows down with age as in solar-type stars,
the rapid rotators in the field must be similar to the Pleiades in age, or
even younger. 
We therefore have a significant population of young low-mass stars
in the solar neighbourhood, generally with low proper motions, and as we
see, quite low radial velocities relative to the Sun. This means that
significant star formation has taken place within a few tens of parsecs of
the Sun in the last few tens or hundred megayears \citep{b00}.

An interesting question is whether the fast rotators have stronger H$\alpha$
emission than the slower rotators, or whether saturation sets in at a
relatively slow rotation speed \citep{jj00}. \citet{ts00}, in a careful
study using much larger telescopes and spectrographs, showed that in the
Pleiades there is a slight positive correlation between rotation speed and
H$\alpha$ equivalent width, but not for $v \sin i > 20 $ km s$^{-1}$, which is our
lower threshold. \citet{rm00} show that $L_{H\alpha}/L_{bol}$ has no
correlation with rotation in either the Hyades or Pleiades, but there does
appear to be a correlation of activity with age, with the Pleiades showing a
factor of ten greater activity. This is extended to other clusters and ages
in Figures 5.19 and 5.20 of \citet{rh00}. 

\begin{figure}
\figurenum{12}
\plotone{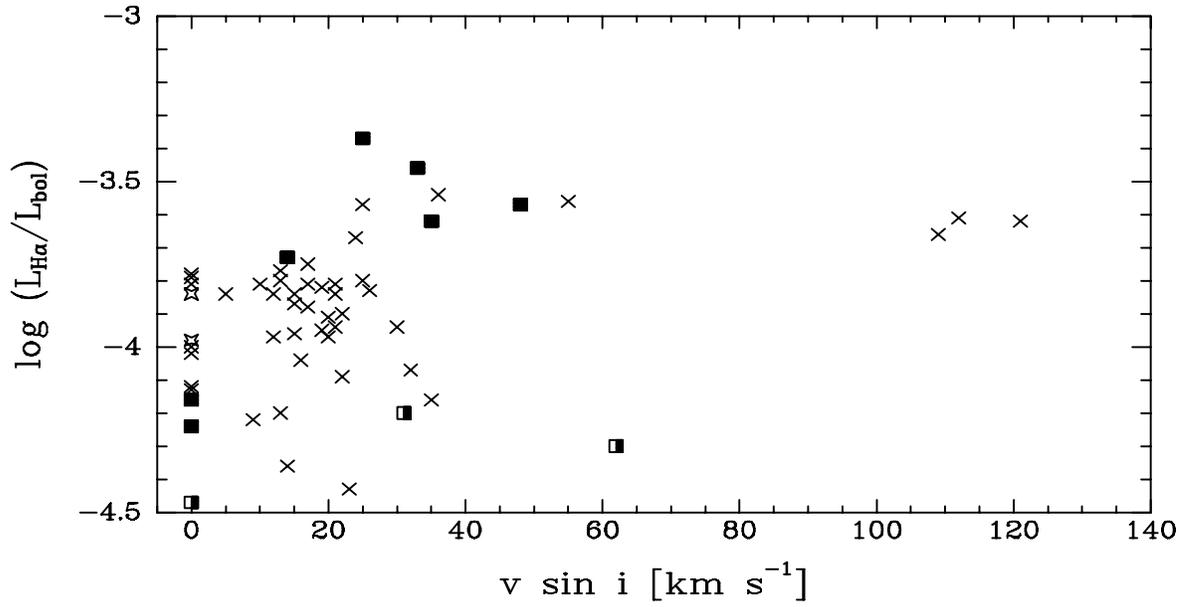}
\figcaption[fig12.eps]{\label{fig12}
Activity strength $L_{H\alpha}/L_{bol}$ versus rotation speed. Plain crosses
represent apparently single stars, the star symbols  represent components
 of double-lined
binaries, while filled squares represent members of visual doubles. The
half-filled squares are suspected spectroscopic binaries.}
\end{figure}

 Figure~\ref{fig12} shows the activity measure $L_{H\alpha}/L_{bol}$ as a
function of $v sin i$ for our sample, averaged for each star. There is at
most a slight positive dependence of H$\alpha$ flux on rotation speed, but
given the wide range of rotation speeds in our sample, it is interesting
that there is so little dependence, if any, on rotation. The ``Skumanich
Law" \citep{sk72}, relating activity to rotation in solar-type dwarfs, does
not seem to apply to mid to early M dwarfs. 

 Our results and those of
\citet{ts00} show that any saturation effect in H$\alpha$ emission must take
effect below $v \sin i = 20 $ km s$^{-1}$. Combined with the X-ray results
of \citet{jj00}, we see that activity indicators in mid to early M dwarfs
are strongly saturated at rotation rates above some speed below $v \sin i =
20 $ km s$^{-1}$; \citet{rh00} infer that the rotation threshold for
activity is below 2 km s$^{-1}$. The results of this paper further confirm
the idea that rotation and activity are not correlated in the mid and earlier M
dwarfs.

The two SB2 dMe systems and the suspected binaries all have normal
H$\alpha$ emission strength. They have low rotation rates (the width of
H$\alpha$ in RX~J0102.4+4101 is probably not caused by rotation). While high
resolution measurements are needed to accurately measure their rotation, it
appears that the emission associated with tidally induced rotation in close
dMe binaries is the same as in single dMe stars.

\begin{figure}
\figurenum{13}
\plotone{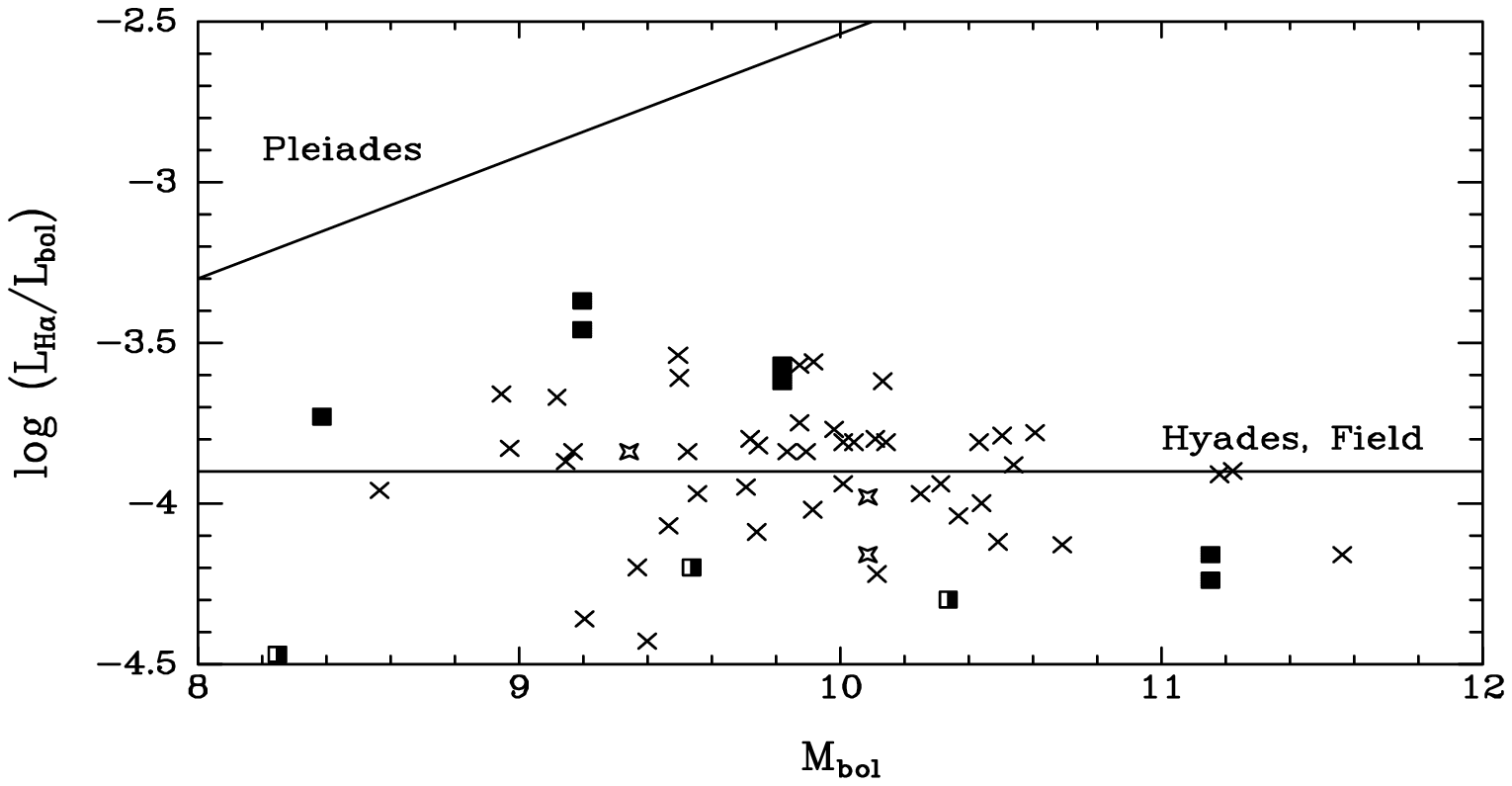}
\figcaption[fig13.eps]{\label{fig13}
Activity strength as a function of absolute bolometric magnitude. Symbols have same
connotation as in Fig.~\ref{fig12}. The {\it mean relations} for the Pleiades, and
the Hyades and field, are also shown. }
\end{figure}

Figure~\ref{fig13} shows activity strength $L_{H\alpha}/L_{bol}$ as a
function of bolometric absolute magnitude M$_bol$. The mean relation for
the Pleiades is estimated from Figure 9 of \citet{rm00}, while the mean
value of -3.9 for the Hyades and the field has been obtained by \citet{rm00}
and \citet{hgr96}, respectively. In this diagram, our sample has the same
distribution as the field and Hyades samples, and lies below the Pleiades
distribution, where the less massive stars appear to be more active. 

\begin{figure}
\figurenum{14}
\plotone{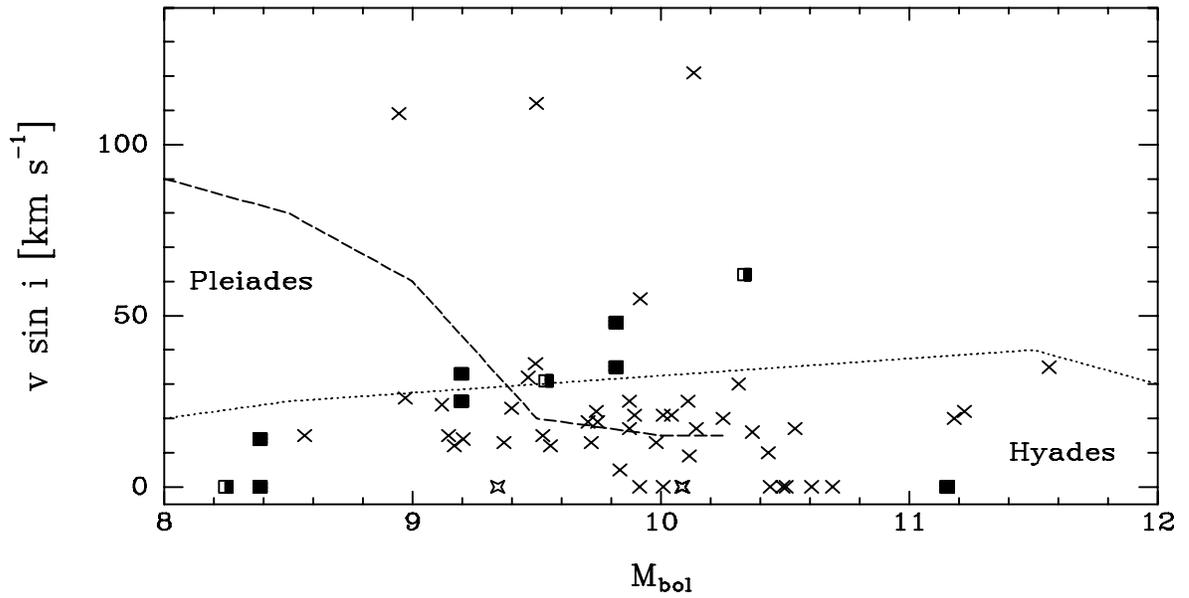}
\figcaption[fig14.eps]{\label{fig14}
Rotation speed as a function of absolute bolometric magnitude. Symbols as in
Fig.~\ref{fig12}. The {\it upper envelopes} of the Pleiades and Hyades
distributions are shown.}
\end{figure}

While the activity distribution as a function of mass looks just like the
field and the Hyades, rotation as a function of mass is somewhat different,
as seen in Figure~\ref{fig14}. Here the lines represent the {\it upper
envelopes} of rotation in
the Hyades and Pleiades. The UFRs of our X-ray selected sample are cooler
(less massive) than the Pleiades, and rotate more rapidly than the Hyades.
This suggests that the stars in our sample either acquired their angular
momentum in a different manner than the M dwarfs in the Pleiades, or that
our rapid rotators are significantly younger than the Pleiades and have not
undergone as much braking. It is also possible that the sources of Pleiades
rotation data as compiled by \citet{rm00} are incomplete at the faint end.
These diagrams convincingly show that for these stars the activity strength
does not depend on rotation speed.

An interesting exercise will be to photometrically monitor the fast and
ultra-fast rotators to obtain their rotation periods. We expect spots and
photometric variability, as well as flares, given the H$\alpha$ variability
we have observed, and the presence of some of our stars in the catalog of
flare stars \citep{gk99}.  Are the true rotation speeds even higher than the
observed $v \sin i$ values? Trigonometric parallaxes combined with
photometry will allow estimates of radii to be made. How much larger are
these stars than ``main sequence" M dwarfs? Are they still contracting to
the main sequence (certainly the case for RX~J1132.7-2651A = TWA 8A), or do
internal magnetic fields suppress convection in the core and thereby
increase the radii of these stars \citep{mm01}? What sort of magnetic
braking takes place in these stars?

\acknowledgements
The authors thank the staff of the David Dunlap Observatory, especially
Shenton Chew, Archie Ridder and the Associate Director, Slavek Rucinski, for
ensuring that all equipment worked well during this long project. Hugh Zhao
cheerfully provided computer support.

SWM (Research Grant) and MDG (Postgraduate Scholarship) gratefully
acknowledge support from the Natural Sciences and Engineering Council of
Canada. SWM acknowledges the hospitality of the Observatories of the
Carnegie Institution of Washington while this paper was completed.
 
The research has made use of the SIMBAD database, operated at the CDS,
Strasbourg, France and accessible through the Canadian Astronomy Data
Centre, which is operated by the Herzberg Institute of Astrophysics,
National Research Council of Canada. The CADC's Digitized Sky Survey
facility was essential to make the acquisition of objects an efficient
operation and we thank the staff of the CADC for their support.

The Digitized Sky Survey was produced at the Space Telescope Science
Institute under U.S. Government grant NAG W-2166. The images of these
surveys are based on photographic data obtained using the Oschin Schmidt
Telescope on Palomar Mountain and the UK Schmidt Telescope. The plates were
processed into the present compressed digital form with the permission of
these institutions.

This publication makes use of data products from the Two Micron All Sky
Survey, which is a joint project of the University of Massachusetts and the
Infrared Processing and Analysis Center/California Institute of Technology,
funded by the National Aeronautics and Space Administration and the National
Science Foundation.

We also thank the IRAF support team at Kitt Peak National Observatory for
their assistance. An anonymous referee made several valuable suggestions
which have improved the conclusions we have reached.

\clearpage
 
\noindent

 

 
 
 



\end{document}